\documentclass[12pt]{article}

\usepackage{slashed}
\usepackage{color, verbatim}
\usepackage{latexsym}
\usepackage{amsmath,accents}
\usepackage{graphicx}

\usepackage{cite}
\usepackage{hyperref}

\usepackage{amssymb}
\usepackage{graphicx}
\usepackage{bm}
\usepackage{hyperref}
\usepackage{adjustbox}

\makeatletter

\@addtoreset{equation}{section}
\makeatother

\begin{document}
\thispagestyle{empty}

\hfill FTUV-17-0508.7937

\hfill IFIC/17-24 

\begin{center}

\begin{center}

\vspace{.5cm}

{\Large\sc Direct Bounds on Heavy Top-Like Quarks \\[0.5cm] 
With Standard and Exotic Decays}

\end{center}

\vspace{0.8cm}

\textbf{
Mikael Chala}\\

\vspace{1.cm}
{\em {Departament de F\'isica T\`eorica, Universitat de Val\`encia and IFIC, Universitat de
Val\`encia-CSIC, Dr. Moliner 50, E-46100 Burjassot (Val\`encia), Spain}}

\end{center}

\begin{abstract}
Heavy vector-like quarks with electric charge $Q=2/3$ (also called \textit{heavy tops}) appear naturally in many extensions of the Standard Model. Although these typically predict the existence of further particles below the TeV scale, direct searches for heavy tops have been performed assuming that they decay only into SM particles.
The aim of this paper is to overcome this situation. We consider the most constraining experimental LHC searches for vector-like quarks, including analyses of the 36 fb$^{-1}$ of data collected in the latest run at 13 TeV of center of mass energy, as well as searches sensitive to heavy tops decaying into a new scalar, $S$. Combining all these, we derive bounds for arbitrary values of the heavy top branching ratios. A simple code that automatizes this process is also provided. At the physics level, we demonstrate that bounds on heavy tops are not inevitably weaker in the presence of new light scalars. We find that heavy tops with masses below $\sim 900$ GeV are excluded by direct searches, independently of whether they decay into $Zt, Ht, Wb$ or $St$ (with $S$ giving either missing energy of bottom quarks) or into any combination of them.
\end{abstract}

\newpage

\tableofcontents

\newpage

\section{Introduction}

Heavy vector-like quarks with electric charge $Q=2/3$ and masses around the TeV scale appear naturally in many extensions of the Standard Model (SM). We will refer to them as \textit{heavy tops} or \textit{top partners} interchangeably.
Since they are colored, they can be produced in pairs via QCD interactions; their signatures depending only on the way they decay. These have been widely studied assuming that these resonances decay only into SM particles~\cite{AguilarSaavedra:2005pv,AguilarSaavedra:2009es,Aguilar-Saavedra:2013qpa}, namely into $Zt$, $Ht$ and $Wb$ (charged-conjugated decays are also understood). However, many models of new physics predict also the existence of further particles, into which the heavy tops could decay. Consequently, these models can not be easily constrained in light of current analyses.

This is the case, for example, of composite Higgs models (CHM)~\cite{Kaplan:1983fs,Kaplan:1983sm,Dimopoulos:1981xc} with partial compositeness~\cite{Kaplan:1991dc}. Apart from the minimal setup, based on the symmetry-breaking pattern $SO(5)/SO(4)$~\cite{Agashe:2004rs}, all realizations of the composite Higgs paradigm contain new scalars of electroweak (EW) mass. In particular singlets, as for example $SO(6)/SO(5)$~\cite{Gripaios:2009pe}, $SO(7)/SO(6)$~\cite{Chala:2016ykx}, $SO(7)/G_2$~\cite{Chala:2012af}, $SO(6)/SO(4)$~\cite{Sanz:2015sua} or $SU(5)/SO(5)$~\cite{Vecchi:2013bja}, among others. Note also that the minimal CHM is in no way preferred over the non-minimal realizations. More the contrary: non-minimal CHMs can provide dark matter candidates~\cite{Frigerio:2012uc,Mrazek:2011iu,Chala:2012af,Fonseca:2015gva,Ma:2017vzm,Ballesteros:2017xeg}, explanations for the observed baryon-anti-baryon asymmetry~\cite{Espinosa:2011eu,Chala:2016ykx} and feasible UV completions~\cite{Caracciolo:2012je,Barnard:2013zea,Ferretti:2014qta,Vecchi:2015fma,Ma:2015gra,Ferretti:2016upr}. Therefore, the top partner phenomenology can be totally different from what current experimental analyses consider. As it has been anticipated in previous references~\cite{Serra:2015xfa, Anandakrishnan:2015yfa}, new decay modes must be also taken into account.

Recent works~\cite{Anandakrishnan:2015yfa,Cacciapaglia:2015eqa,Fan:2015sza,Banerjee:2016wls,Niehoff:2016zso} have made a first attempt to address this question by exploring the signatures of new channels. However, this approach is valid \textit{only} when new decays are dominant and hence of little help.  Likewise, standard analyses apply \textit{only} when the top-partners decay mostly into SM particles, namely into $Ht, Zt$ and $Wb$. Overall, top-partners with both standard and exotic decays can be bounded by no means in regard of present analyses. Our aim in what follows is to fill this gap by extending previous efforts in two ways. First, we consider all possible sets of branching ratios for the top partners, counting not only new channels but also elusive decays, \textit{i.e.} those that evade current searches. They have the sole (but so far rather unexplored) effect of making the branching ratios of all observable topologies not add to 1 (see ref.~\cite{Barducci:2014ila} for previous considerations of this possibility). And second, we include LHC data acquired at both $8$~\cite{Aad:2015kqa} and $13$~\cite{ATLAS-CONF-2016-102,ATLAS-CONF-2016-104,ATLAS-CONF-2017-015,CMS:2016hxa} TeV of center of mass energy (\textit{c.m.e.}). All results are given in terms of tables and plots than can be trivially used to constrain generic top-like resonances. To simplify further this task, we also provide a very simple code that implements our findings. It can be found in \url{http://github.com/mikaelchala/vlqlimits}. We intend to update this program with the inclusion of incoming analyses and other signatures of top and also bottom partners.  

Evidently, heavy tops can decay into several new channels without conflicting with the SM gauge symmetries. We do not explore all of them in this article, nor are they included in the mentioned code in its current form. Instead, we restrict ourselves to the standard channels plus the top-partners decaying into a neutral scalar and a top quark. We assume that the former either decays into bottom quarks or it escapes detectors. As a matter of fact, these production modes are expected to dominate over others in concrete CHMs. This is discussed in section~\ref{sec:assumptions}. The rest of the article is structured as follows. In section~\ref{sec:smdecays} we describe the status of experimental searches that bound the top-partner decays into SM particles. We also explain how these can be combined without necessarily recasting the corresponding analyses.  In section~\ref{sec:exotic} we concentrate on searches for heavy tops with exotic decays. Different analyses are recast and applied to simulated events in these topologies and their expected collider signatures are obtained. In section~\ref{sec:final} we describe how to combine all previous outcomes to bound heavy tops of several masses. Plots for a broad set of branching ratios are provided. It is shown that bounds on heavy tops are not inevitably weaker in the presence of new light scalars. Finally, in appendix~\ref{app:software} we comment on how to obtain the code that implements the aforementioned results.

\section{Assumptions}
\label{sec:assumptions}

Inspired by non-minimal CHMs, we assume that the relevant degrees of freedom at the TeV scale involve the SM particles, some new fermionic resonances and at least one real neutral scalar, $S$. The latter, as well as the Higgs, $H$, are supposed to be composite objects resulting from the confinement of a new strongly interacting sector at some scale $f\sim$ TeV. We will denote by $m_H$, $m_S$ and $M$ the masses of $H$, $S$ and the fermionic resonances, respectively. The phenomenological study presented in this article is based on the following assumptions.

%\begin{itemize}
\textbf{Assumption 1}: \textit{The top partners, $T$, are among the lightest fermionic resonances}. This has been explained at length in the CHM literature. A sketch of the argument reads as follows. According to the partial compositeness paradigm~\cite{Kaplan:1991dc}, the elementary fermions, $q$, mix with vector-like resonances, $Q$, through the mixing Lagrangian $\sim \Delta \overline{q}Q+\text{h.c.}$ The latter fermions are the only which, in turn, couple to the Higgs boson, because they are all composite. In the physical basis, however, the massless particles (to be identified with the SM fermions) obtain Yukawa interactions of the order $\sim (\Delta/M)^2$. The smallest $M$ is then that of the fermionic resonance coupling to the SM fermion with the largest Yukawa. Namely the top quark, as we claim. More quantitative analyses can be found, for example, in refs.~\cite{Contino:2006qr,Matsedonskyi:2012ym,Redi:2012ha,Marzocca:2012zn,Pomarol:2012qf,Panico:2012uw}.
 
\textbf{Assumption 2}: \textit{$m_S > m_H$, being in general of a few hundreds GeV}. The scalar potential for $H$ and $S$ can be estimated to be~\cite{Chala:2017sjk}
\begin{equation}
V \sim \frac{3}{(4\pi)^2} f^2 M^2 \left[-\alpha \frac{H^2}{f^2} + \beta \frac{H^4}{f^4} + \gamma \frac{S^2}{f^2} \right] + ...
\end{equation}
where $\alpha,\beta$ and $\gamma$ are dimensionless coefficients of order $\mathcal{O}(1)$, and the ellipsis stand for terms with higher powers of $S$.
Experimental limits from Higgs searches~\cite{Khachatryan:2016vau} and EW precision data~\cite{Ghosh:2015wiz} impose $f \gtrsim 900 $ GeV, while natural arguments suggest that this can not be much larger. Thus, let us take $M\sim f\sim 1$ TeV. It turns out that $m_S$ ranges from $\sim 150$ GeV (for $\gamma \sim 0.5$) to $\sim 200$ GeV (for $\gamma\sim 1$). 
 We can further strengthen this conclusion by focusing on particular models with computable $m_S$. Among these, we find CHMs in which $S$ is stable. The first such a model we consider is based on the coset $SO(6)/SO(5)$, with the third-generation $q_L$ and $t_R$ mixing with composite resonances transforming in the $\mathbf{20}$ and the $\mathbf{1}$ representations of $SO(6)$, respectively. The scalar potential at the leading order adopts the form~\cite{Serra:2015xfa,Chala:2017sjk}
 \begin{equation}
  V = f^2\bigg[c_1 - \frac{7}{4}c_2\bigg] H^2 + (c_2 - c_1) H^4 - c_2 f^2 S^2 + \cdots
 \end{equation}
 where $c_1$ and $c_2$ are the only free parameters not constrained by the symmetries. These can be traded by the known values of the Higgs mass term, $\mu_H^2$, and the quartic coupling, $\lambda_H\sim 0.13$:
 \begin{equation}
  V = \frac{1}{2}\mu_H^2 H^2 + \frac{1}{4}\lambda_H H^4 + \frac{1}{3}f^2 \left[1-2\frac{v^2}{f^2}\right] \lambda_H S^2 + \cdots
 \end{equation}
 with $v\sim 246$ GeV the EW VEV. If we take $f\sim 1$ TeV, we obtain $m_S\sim 300$ GeV. The second model we consider is based on the larger coset $SO(7)/G_2$, with the third generation $q_L$ and $t_R$ mixing with the $\mathbf{35}$ and the $\mathbf{1}$ representations of $SO(7)$, respectively. The scalar potential can then be written as~\cite{Ballesteros:2017xeg}
\begin{equation}
 V = \mu_H^2 H^2 + \lambda_H H^4 + \frac{1}{3} f^2 \bigg[1-\frac{8}{3}\frac{v^2}{f^2}\bigg]\lambda_H S^2 + \cdots
\end{equation}
 Again, for $f \sim 1$ TeV  we obtain $m_S \sim 300$ GeV.
 
\textbf{Assumption 3}: \textit{If $S$ is not stable, it decays mostly into bottom quarks}. $S$ is a composite particle, and therefore interacts stronger with the heavier particles. Thus, with tops and bottoms. However, in light of the discussion above, it is clear that there exists a natural regime in which $S$ has no kinematics space to decay into $t\bar{t}$, nor into $HH$. Besides, the latter can be further suppressed if $S$ is CP odd. 

As a consequence of these three assumptions, the heavy tops can only decay sizably into $Zt, Ht, Wb$ and $St$ (with $S$ either decaying into bottom quarks or escaping detection). With this in mind, in next sections we obtain bounds on $M$ for arbitrary branching ratios and for different
values of $m_S$ in between 100 and 400 GeV. We focus on the pair production channel, because it is driven by model-independent QCD interactions. In this respect, however, two comments are in order. \textit{(i)} In principle, spin-1 resonances, as for example heavy gluons, could also mediate this process~\cite{Chala:2014mma,Araque:2015cna,Azatov:2015xqa}. Were this the case, the production cross section would be larger: our bounds considering only the QCD contribution would be conservative. \textit{(ii)} Likewise, in the context of CHMs, the cross section for single production of heavy tops might be also large~\cite{Matsedonskyi:2014mna,Matsedonskyi:2015dns}.

\section{Searches for vector-like quarks with standard decays}
\label{sec:smdecays}
 \begin{figure}[t]
\begin{center}
  \includegraphics[width=0.49\columnwidth]{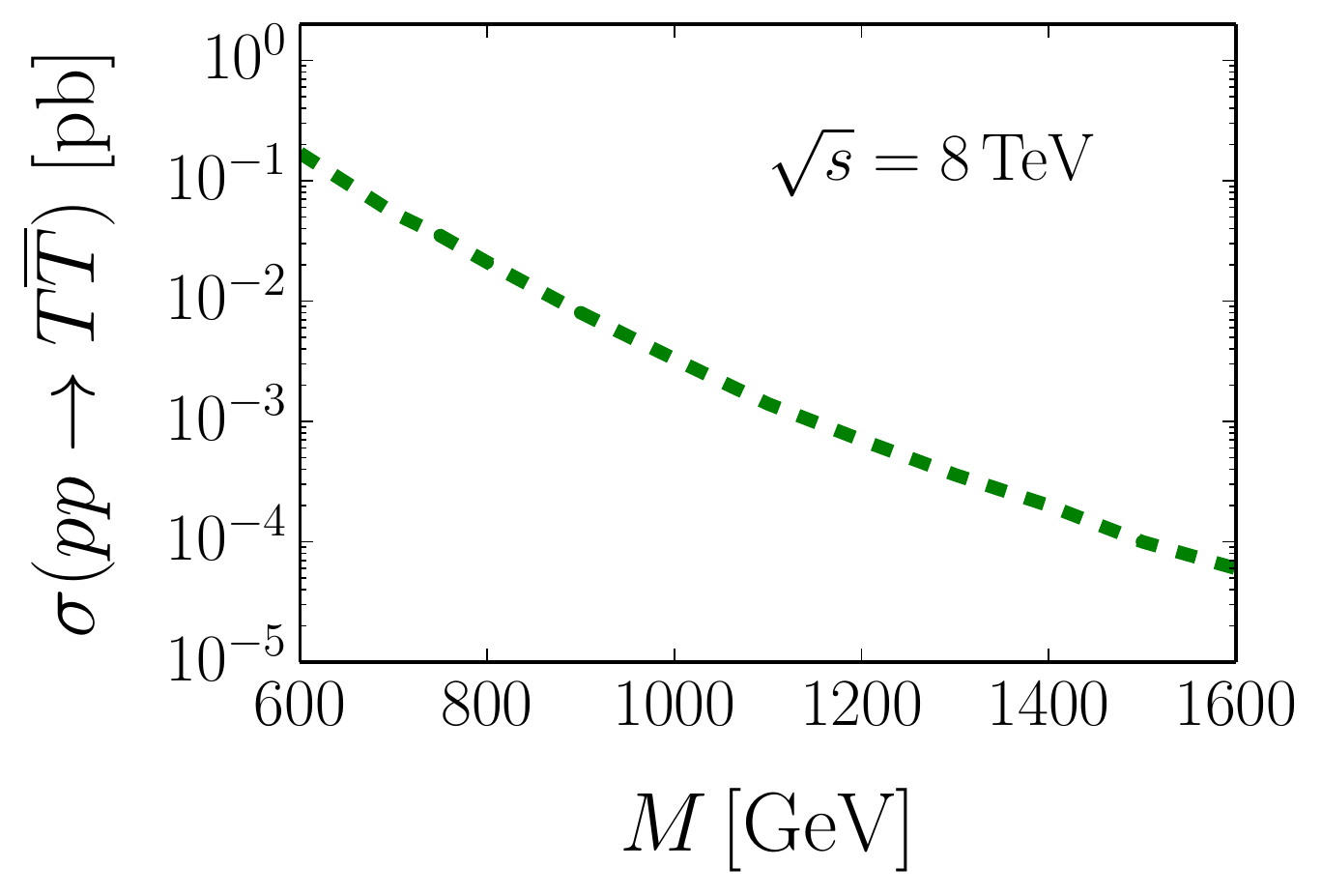}
  \includegraphics[width=0.49\columnwidth]{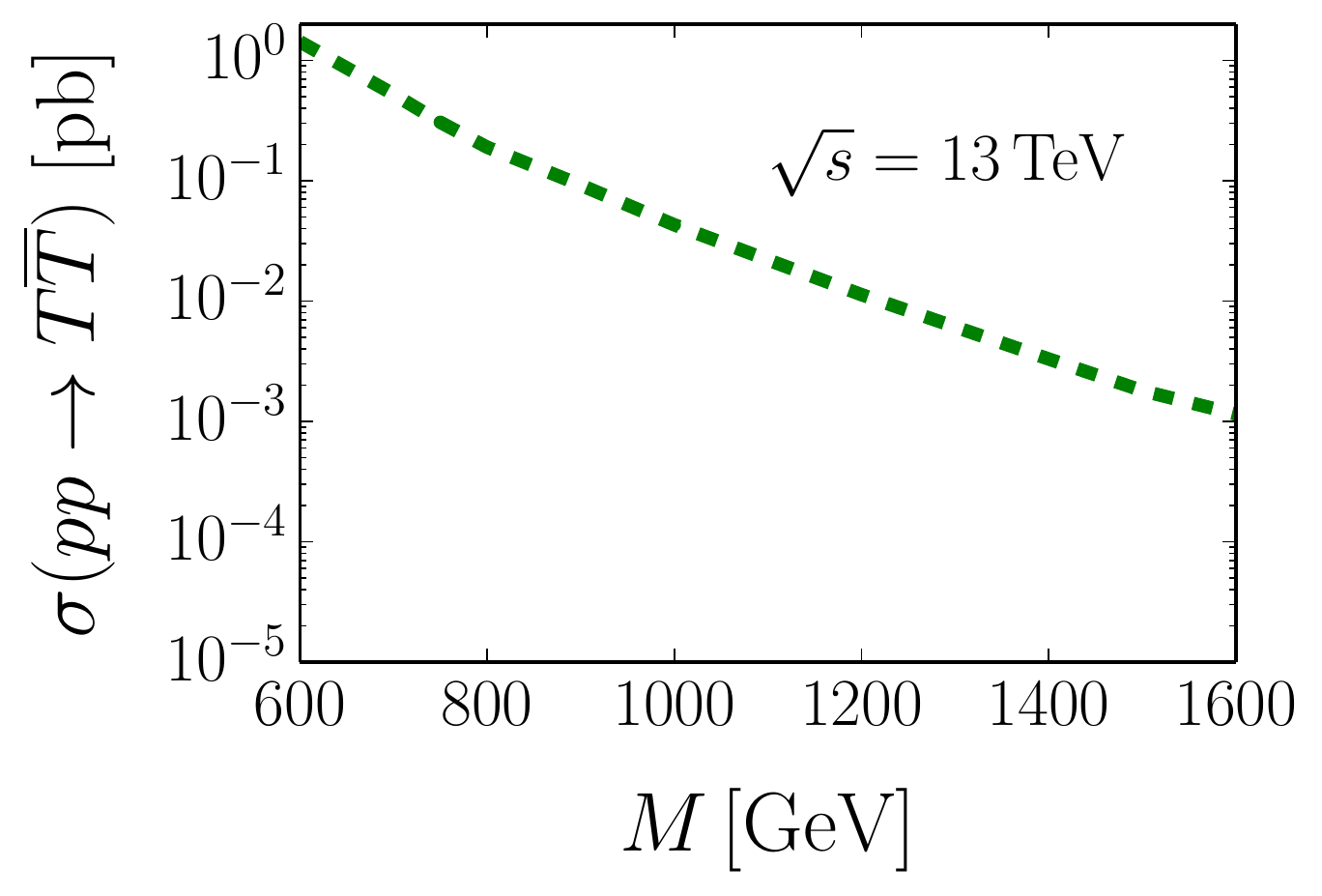}
\end{center}
\caption{\it NNLO cross sections for pair production of top partners~\cite{Aad:2015kqa,ATLAS-CONF-2016-104} at 8 (left) and 13 (right) TeV of c.m.e.}\label{fig:xsecs}
\end{figure}
Standard searches for QCD pair-produced top partners assume that these decay \textit{only} into SM particles, namely
\begin{equation}\label{eq:brs}
 \text{BR}(T\rightarrow Ht) +  \text{BR}(T\rightarrow Wb) + \text{BR}(T\rightarrow Zt) = 1~.
\end{equation}
In fact, bounds are typically given by plots showing excluded regions in the plane of the first two branching ratios (see for example all searches discussed in ref.~\cite{Araque:2016jrb}). 
Consequently, bounds for top partners with elusive decays, for which this equality does not hold, cannot be read from current plots. Some constraints can be still obtained by rescaling the ones given for each channel separately. However, these disregard all signal events in which the two heavy tops decay into different particles. Bounds obtained in this way are hence highly underestimated. At first glance, any attempt of departing from eq.~\ref{eq:brs} requires reinterpreting the data from scratch (\textit{e.g.} recasting the corresponding analyses).

Fortunately, this exercise can be greatly simplified in a counting experiment if limits are provided for at least three different sets of branching ratios. We will illustrate how to proceed in that case in the first SM channel to which we pay attention, namely $T\rightarrow Zt$.

\subsection{$T\rightarrow Zt$}
\label{sec:zt}
Let us consider a counting experiment for $T\overline{T}\rightarrow Zt + X$, with $X$ being any standard $T$ decay channel. Let us assume that bounds on the QCD model-independent pair-production cross section are known for $\text{BR}(T\rightarrow Zt) = 1$. We will dub this quantity $\sigma_{1}$. Then, it is clear that the maximum amount of allowed signal events after cuts, $N$, can be written as $N = \sigma_1\, L \, \epsilon(Zt Zt)$,
where $L$ stands for the integrated luminosity considered in the analysis and $\epsilon(Zt Zt)$ is the efficiency for selecting signal events in the $Zt + Zt$ channel.

Let us further assume that limits for the cases $\text{BR}(T\rightarrow Ht) = \text{BR}(T\rightarrow Zt) = 0.5$ (the \textit{doublet} case) and $2~\text{BR}(T\rightarrow Ht) = 2~\text{BR}(T\rightarrow Zt) = \text{BR}(T\rightarrow Wb) = 0.5$ (the \textit{singlet} case) are also known. We will name their values by $\sigma_D$ and $\sigma_S$, respectively. Then, we get that the following equations approximately hold:
\begin{equation}
 N \sim \sigma_D\, L\, \bigg[\frac{1}{4} \epsilon(ZtZt) + \frac{1}{2}\epsilon(ZtHt)\bigg] \sim \sigma_S\, L\, \bigg[\frac{1}{16} \epsilon(ZtZt) + \frac{1}{8}\epsilon(ZtHt) + \frac{1}{4}\epsilon(ZtWb)\bigg]~.
\end{equation}
We have (conservatively) neglected $\epsilon(Ht Ht), \epsilon(Wb Wb)$ as well as $\epsilon(Ht Wb)$, given that the experimental search tags mainly $Zt+X$ events. From the equation above, all the relevant efficiencies can be estimated, and therefore the number of expected signal events passing all the analysis cuts, $N_\text{sig}$, for arbitrary values of the branching ratios:
\begin{align}\label{eq:signal}\nonumber
 N_\text{sig} \sim \sigma \, L \,\bigg[\text{BR}(T\rightarrow Zt)^2\epsilon(Zt Zt) &+ 2~\text{BR}(T\rightarrow Zt)\,\text{BR}(T\rightarrow Ht)~\epsilon(Zt Ht) \\
 & + 2~\text{BR}(T\rightarrow Zt)\,\text{BR}(T\rightarrow Wb)~\epsilon(Zt Wb) \bigg]~.
\end{align}
In this expression, $\sigma$ represents the theoretical cross section for pair-production of top partners. This is shown in fig.~\ref{fig:xsecs}, assuming that is driven by QCD interactions.
The number of signal events can be compared with the number of observed and predicted SM events. Altogether, this method can be used to obtain bounds on top-like resonances decaying into SM particles even when eq.~\ref{eq:brs} is not fulfilled. Conversely, if eq.~\ref{eq:brs} is satisfied, the constraints obtained using this approach are slightly conservative. Let us apply this procedure to a real example. This will allow us to check the goodness of the method. In addition, the results obtained in this way will be used in next parts of this article, when bounds combining different decay modes (including exotic $T$ decays) will be given.
\begin{table}[t]
 \begin{center}
\begin{adjustbox}{width=0.95\textwidth}
\footnotesize
\begin{tabular}{||l|c|c|c||}\hline
efficiency (\%) & $M = 800$ GeV & $M = 1000$ GeV & $M = 1100$ GeV\\  
\hline
 $Zt + Zt$ & 0.5 & 1.0 & 1.3  \\
 $Zt + Ht$ & 0.3 & 0.8 & 0.9  \\
 $Zt + Wb$ & 0.2 & 0.4 & 0.5  \\
 \hline
 \end{tabular}
 \end{adjustbox}
 \end{center}
 \caption{\it Relevant efficiencies of the analysis of ref.~\cite{ATLAS-CONF-2017-015} for three different masses of the heavy top as obtained using the procedure outlined in the text.}\label{tab:efficiencies}
\end{table}

\begin{figure}[t]
\begin{center}
  \includegraphics[width=0.32\columnwidth]{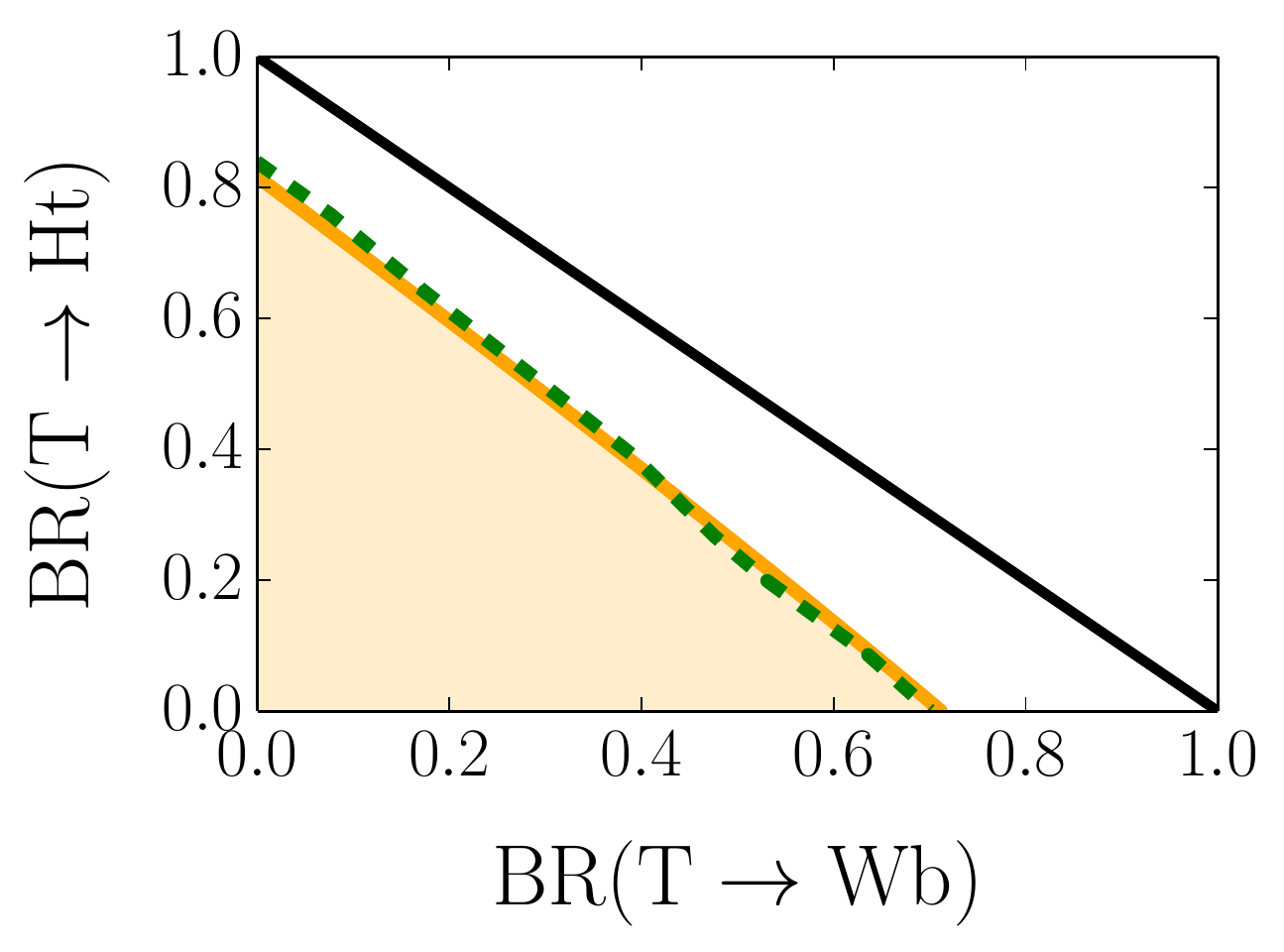}
  \includegraphics[width=0.32\columnwidth]{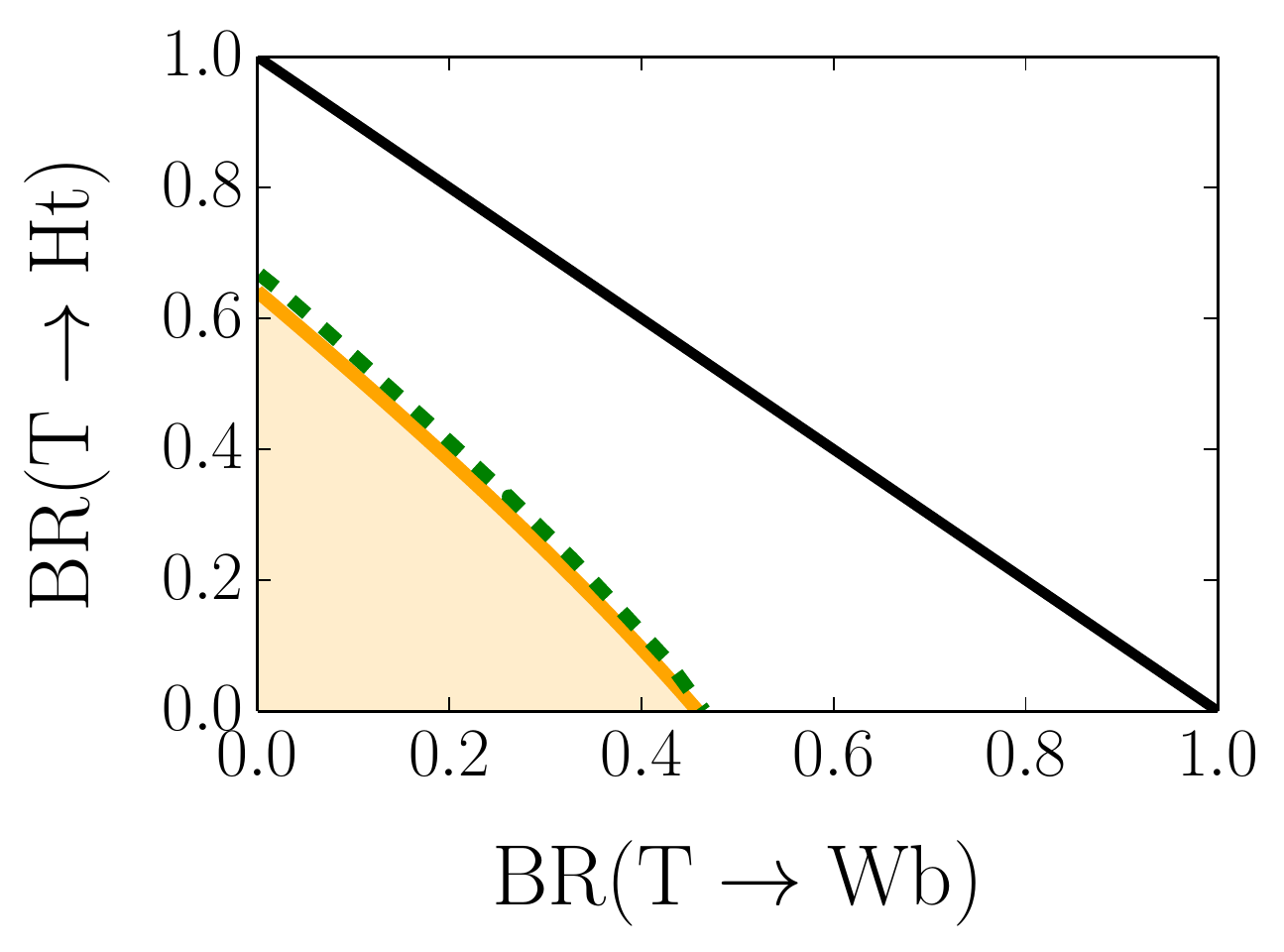}
  \includegraphics[width=0.32\columnwidth]{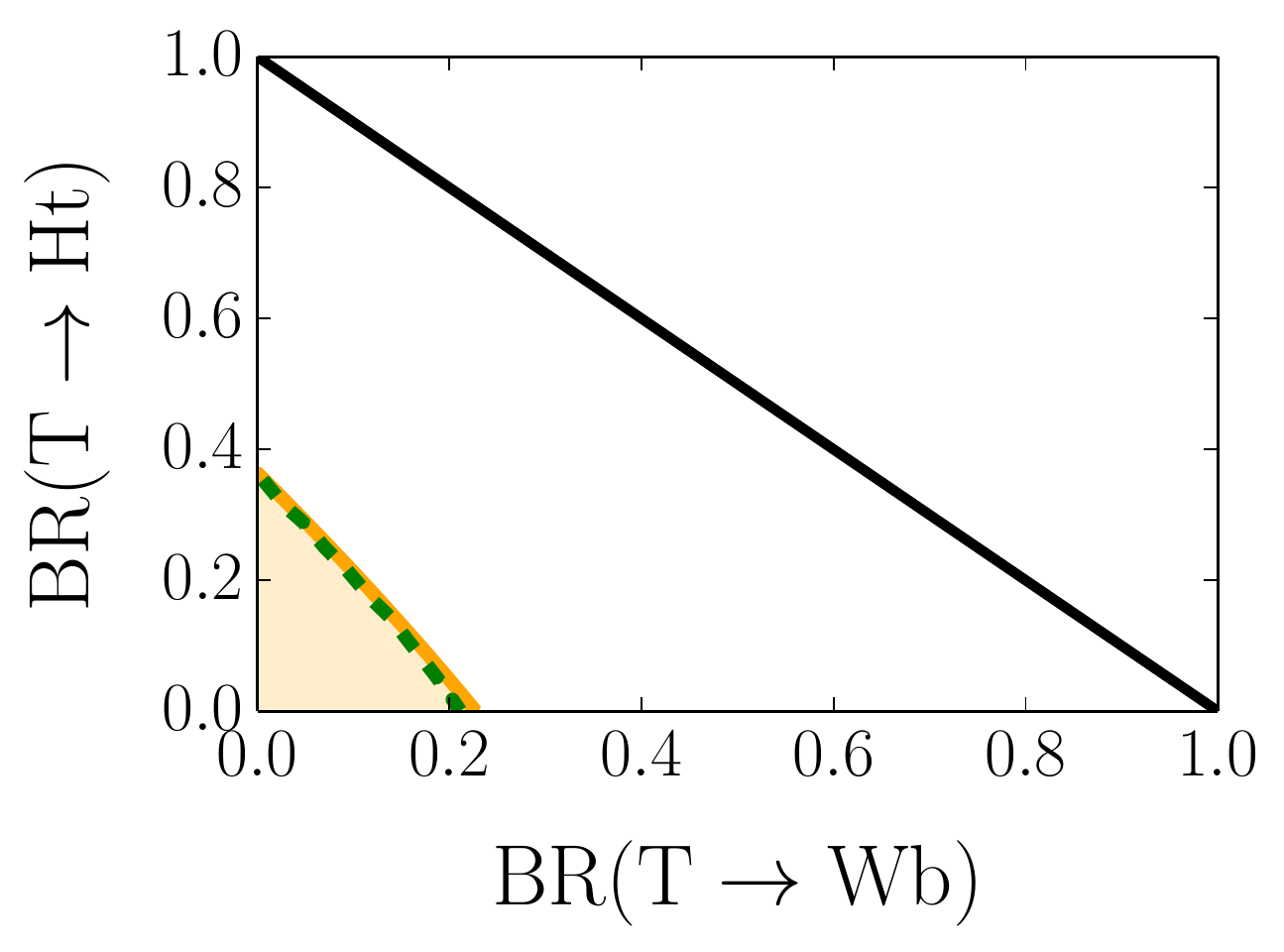}
\end{center}
\caption{\it Left) Region of the plane of branching ratios excluded by the $Zt+X$ analysis of ref.~\cite{ATLAS-CONF-2017-015} for $M = 800$ GeV. The dashed green lines represent the constraints provided by the experimental collaboration, while the orange area is the one we obtain with the simplified method outlined in the text. Middle) Same as Left) but for $M = 1000$ GeV. Right) Same as Left) but for $M=1100$ GeV.}\label{fig:valZt}
\end{figure}

We consider the latest search for $Zt+X$ of ref.~\cite{ATLAS-CONF-2017-015}. This is a counting experiment based on 36 fb$^{-1}$ of collected data at 13 TeV of c.m.e. at the ATLAS experiment. It focuses mainly on the invisible decay of the $Z$ boson. Among the most important cuts, the analysis requires the presence of exactly one light lepton (either electron or muon), missing transverse energy $E_T^{\text{miss}} > 350$ GeV, the presence of at least four hard small-radius jets and the presence of two large-radius jets with high $p_T$. The search is optimized for a heavy top with mass $M\gtrsim 1$ TeV, for which the selection efficiency is reported to be $\sim 1 \%$. (Note that this small efficiency takes into account all SM decay modes.) Bounds at the 95 \% C.L. for the case $\text{BR}(T\rightarrow Zt) = 1$ as well as for the singlet and the doublet cases are provided in fig. 6 of that reference for masses between 600 and 1400 GeV. Thus, following the procedure outlined above, we obtain the relevant efficiencies. These are written in tab.~\ref{tab:efficiencies} for three representative masses. Clearly, although the efficiencies in the $Zt+Zt$ channel are larger, the efficiencies in the mixed channels are not negligible. The number of predicted SM events passing all cuts is reported to be $N_{\text{SM}} \sim 6.5$, while the number of observed events is $N_{\text{O}} = 7$. The maximum number of allowed signal events can then be obtained using a simple CL$_s$ method~\cite{Read:2002hq}. This method considers the statistic
\begin{equation}
Q = \prod_i \frac{(s_i+b_i)^{\tilde{n}_i}e^{-(s_i+b_i)}}{b_i^{\tilde{n}_i}e^{-b_i}} = e^{-\sum_i s_i}\prod_i\left[1+\frac{s_i}{b_i}\right]^{\tilde{n}_i} = e^{-N_{\text{sig}}} \left[1 + \frac{N_\text{sig}}{N_{\text{SM}}}\right]^{\tilde{N}_{\text{O}}}~.
\end{equation}
In the last equality we have used that the number of independent signal regions, over which the index $i$ runs, is exactly 1 in a counting experiment. Besides, $\tilde{N}_{\text{O}}$ represents a random Poisson distributed variable with mean $N_{\text{sig}}+N_{\text{SM}}$ in the signal+background hypothesis, and with mean $N_{\text{O}}$ in the background-only hypothesis. Let us call $P_{\text{sig+SM}}(Q)$ and $P_{\text{SM}}(Q)$ the random distribution followed by $Q$ in the first and second cases, respectively. Two confidence estimators are defined in the CL$_s$ method:
\begin{equation}
 \text{CL}_{s+b} = 1-\int_{Q_\text{obs}}^\infty P_\text{sig+SM}(Q) dQ, \qquad \text{CL}_b = 1-\int_{Q_\text{obs}}^\infty P_\text{SM}(Q) dQ~,
\end{equation}
where  $Q_{\text{obs}}$ is the value of $Q$ when $\tilde{N}_{\text{O}} = N_\text{O}$. A signal is said to be excluded at the 95 \% C.L. if CL$_{s}=\text{CL}_{s+b}/\text{CL}_s < 0.05$. In this case, this occurs for $N_\text{sig} > 8$. Fixing eq.~\ref{eq:brs}, we compute this ratio using the \texttt{TLimit} class of \texttt{ROOT}~\cite{Brun:1997pa} for the three values of $M$ in tab.~\ref{tab:efficiencies} and for all values of $\text{BR}(T\rightarrow Ht)$ and $\text{BR}(T\rightarrow Wb)$ in the range $(0, 1)$ with steps of $0.01$. $N_\text{sig}$ is computed using eq.~\ref{eq:signal} and the efficiencies quoted in the table. The excluded regions are shown in orange in the three panels of fig.~\ref{fig:valZt}. The bounds obtained in (fig. 7 of) the experimental reference are superimposed with green dashed lines. Clearly, the agreement is excellent. 

\subsection{$T\rightarrow Ht$}
\label{sec:ht}

We follow the same approach as before to obtain the efficiencies for selecting $Ht+X$ events. We consider the experimental analysis of ref.~\cite{Aad:2015kqa}. It is based on 20 fb$^{-1}$ of collected luminosity at 8 TeV of c.m.e. at the ATLAS experiment. Despite the bounds reported by the paper are based on a likelihood analysis of several signal regions, the sensitivity of the search is driven mainly by one of them. So, it can be approximated by a counting experiment. In this signal region, at least six light jets have to be present, of which at least four must be $b$-tagged. Again, exactly one light lepton is required. The analysis reports the observation of 84 events, while 81 were expected from the SM alone. Consequently, no more than 22 signal events are allowed. To validate the approach in this case, we consider a heavy top mass of 750 GeV. This choice is intended to combine the $Ht+X$ channel with the $Wb+X$ one, for which ref.~\cite{Aad:2015kqa} provides bounds too. (For other masses, the regions excluded combining the $Ht+X$ and $Wb+X$ channels are just the trivial overlapping of those excluded by each search separately. We elaborate on this point in the next section.) The bounds that we obtain are shown in the left panel of fig.~\ref{fig:val}. The agreement is worse than in the $Zt+X$ case, the reason being that this analysis is not exactly a counting experiment. The results, however, are conservative.

On top of this analysis, we consider the latest ATLAS search for $Ht+X$ described in ref.~\cite{ATLAS-CONF-2016-104}. It is based on 13.2 fb$^{-1}$ of collected luminosity at 13 TeV of c.m.e. We consider the channel that requires the presence of exactly one light lepton. The search further imposes a stringent cut on the multiplicity of jets and $b$-tagged jets, as well as on the scalar sum of the transverse momenta of all final state objects. Boosted techniques are used to reconstruct hadronically-decaying resonances giving rise to large-radius jets. Contrary to the previous searches, the bounds obtained in this one rely on the several bins of the effective mass, defined as the scalar sum of the transverse momenta of the lepton, the selected jets and the $E_T^\text{miss}$. Consequently, we can not proceed as before. Instead, we consider just the reported excluded cross sections, $\sigma_\text{excl}$, for $\text{BR}(T\rightarrow Ht) = 1$. Thus, a heavy top of a given mass $M$ is ruled out by this analysis if
\begin{equation}
 \sigma(pp\rightarrow T\overline{T})\times \text{BR}(T\rightarrow Ht)^2 > \sigma_\text{excl}~.
\end{equation}
Bounds obtained in this way can not be combined with the ones before, but they will be superimposed to them. As a matter of fact, when this branching ratio is close to 1, this search sets the most stringent constraint.

\subsection{$T\rightarrow Wb$}
\label{sec:wb}

\begin{figure}[t]
\begin{center}
  \includegraphics[width=0.32\columnwidth]{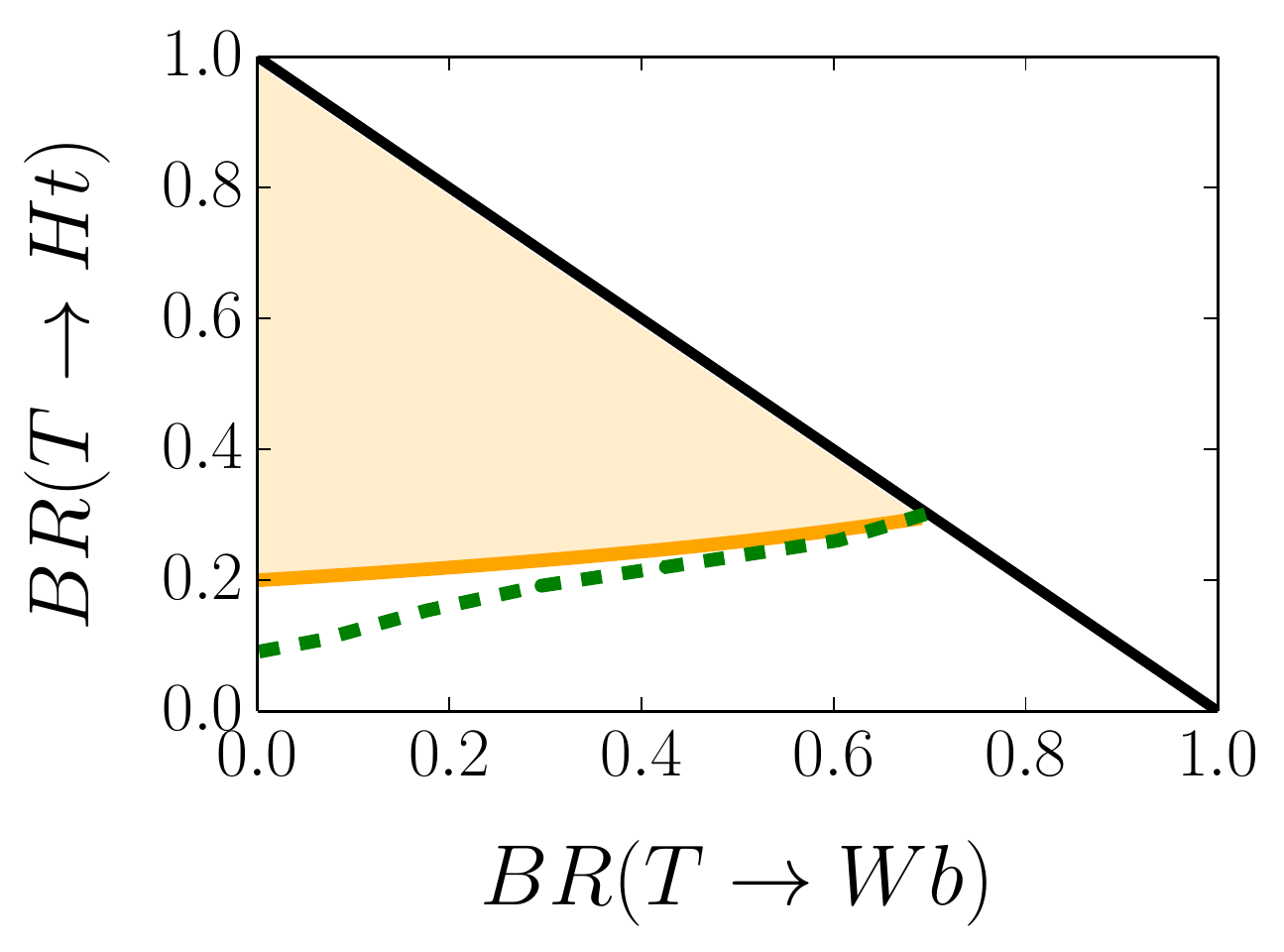}
  \includegraphics[width=0.32\columnwidth]{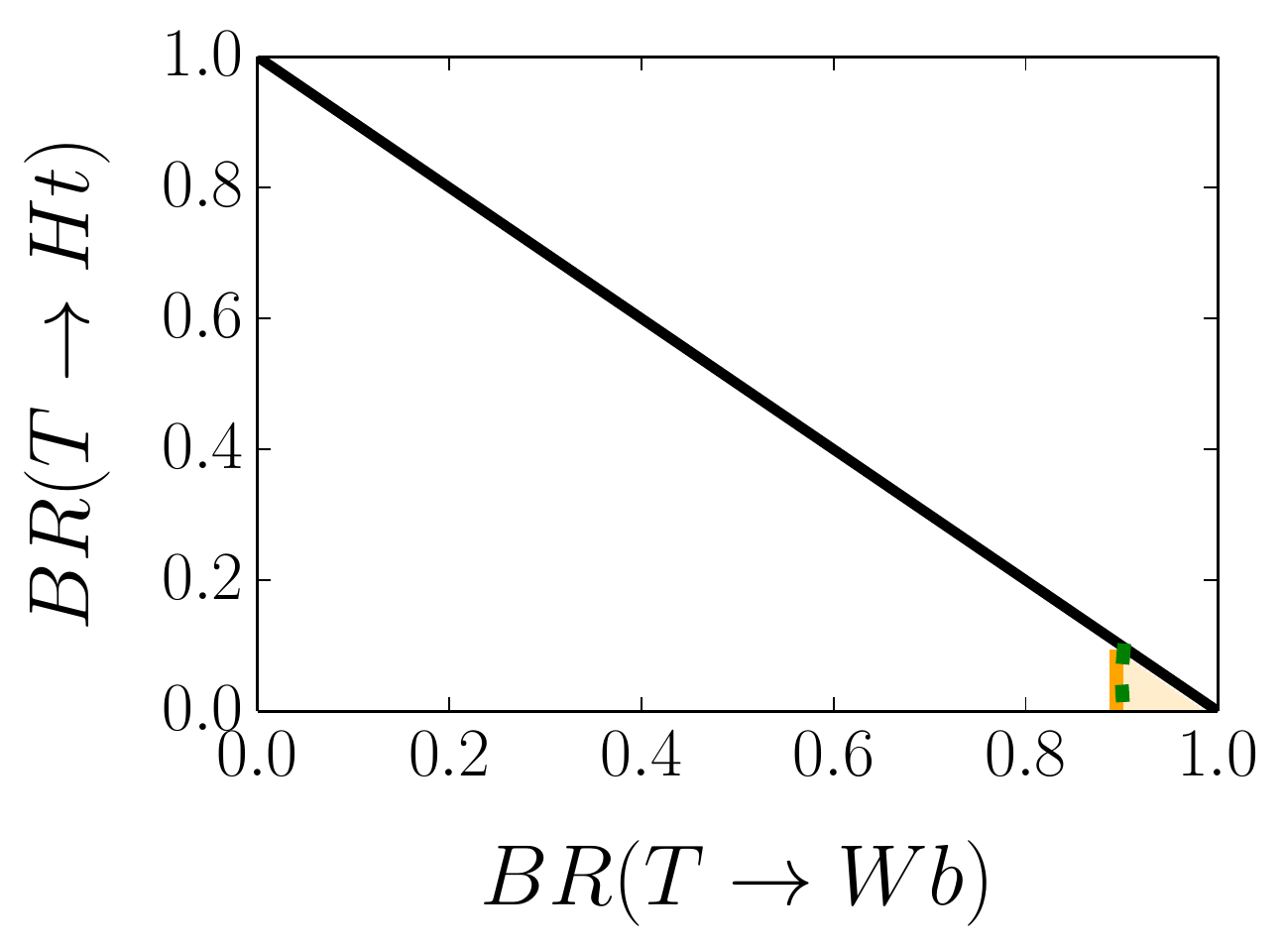}
  \includegraphics[width=0.32\columnwidth]{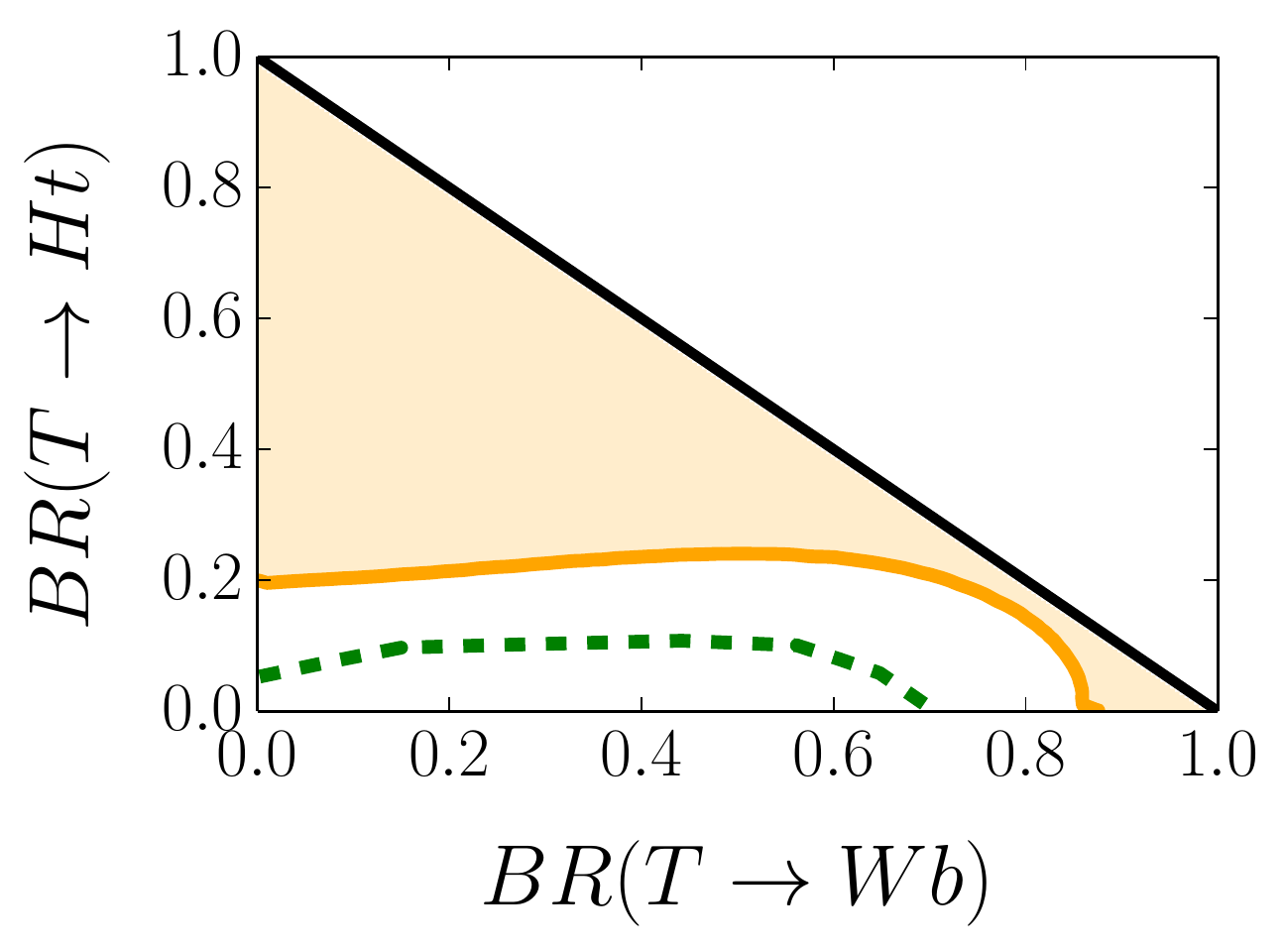}
\end{center}
\caption{\it Left) Region of the plane of branching ratios excluded by the $Ht+X$ analysis of ref.~\cite{Aad:2015kqa} for $M = 750$ GeV. The dashed green lines represent the constraints provided by the experimental collaboration, while the orange area is the one we obtain with the simplified method outlined in the text. Middle) same as Left) but for the $Wb+X$ analysis. Right) Same as Left) but for the combination of both analyses.}\label{fig:val}
\end{figure}

A search for $Wb+X$ has been also carried out in ref.~\cite{Aad:2015kqa}. Among the most important cuts, exactly one light lepton is required, as well as at least four light jets, one of which must be $b$-tagged. Besides, one jet with $p_T > 400$ GeV has to be present, too. This jet results typically from the hadronic decay of one $W$ boson. According to the study, 27.6 SM events are expected to pass all cuts, while 30 have been observed. This implies that no more than 15 signal events are allowed. The analysis does not provide bounds for the doublet case, so we only obtain the efficiency for selecting $Wb +Wb$ events using the simplified method explained above. The region excluded using this approach for $M = 750$ GeV is shown in the middle panel of fig.~\ref{fig:val}. It matches perfectly the one obtained by the experimental collaboration, which means that very few events result from the disregarded $Wb+Ht$ or $Wb+Zt$ channels.

Using the CL$_s$ method described at the beginning of this section, we obtain bounds combining the signal region of the 8 TeV $Ht+X$ analysis with that of the $Wb+X$ one (they are statistically independent). The result is shown in the right panel of the figure. Such a combination has been also provided by the experimental collaboration, and it is depicted by the green-dashed line. Our results are again conservative. On top of this, we also consider the 
latest search for $Wb+X$ described in ref.~\cite{ATLAS-CONF-2016-102}, based on 14.7 fb$^{-1}$ of collected luminosity at 13 TeV of c.m.e. In the final plots, we will also superimpose constraints obtained from those reported for $\text{BR}(T\rightarrow Wb) = 1$.

\section{Searches for vector-like quarks with exotic decays}
\label{sec:exotic}
\subsection{$T\rightarrow St, S\rightarrow b\overline{b}$}
\label{sec:bs}
\begin{figure}[ht]
\begin{center}
  \includegraphics[width=0.49\columnwidth]{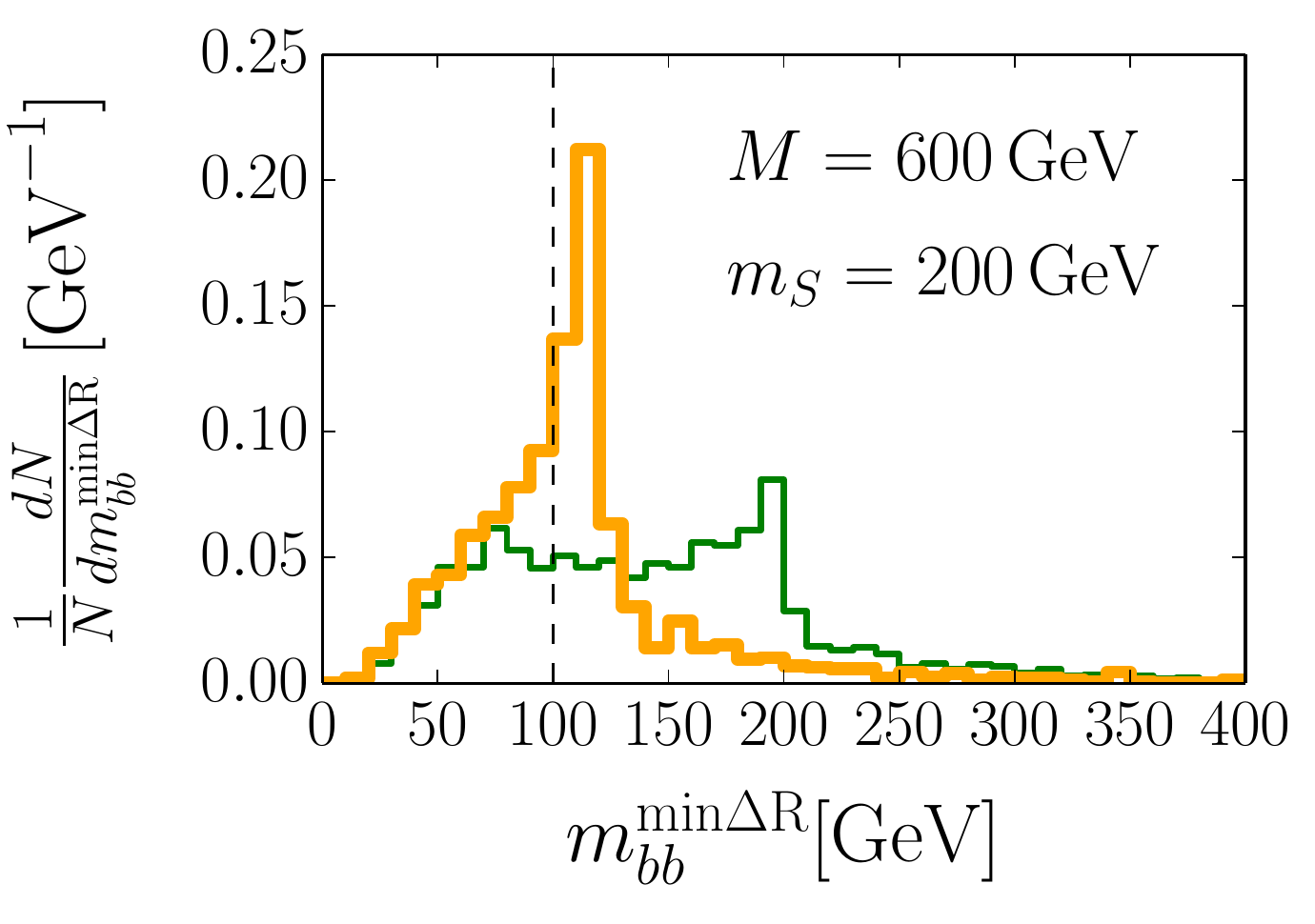}
  \includegraphics[width=0.49\columnwidth]{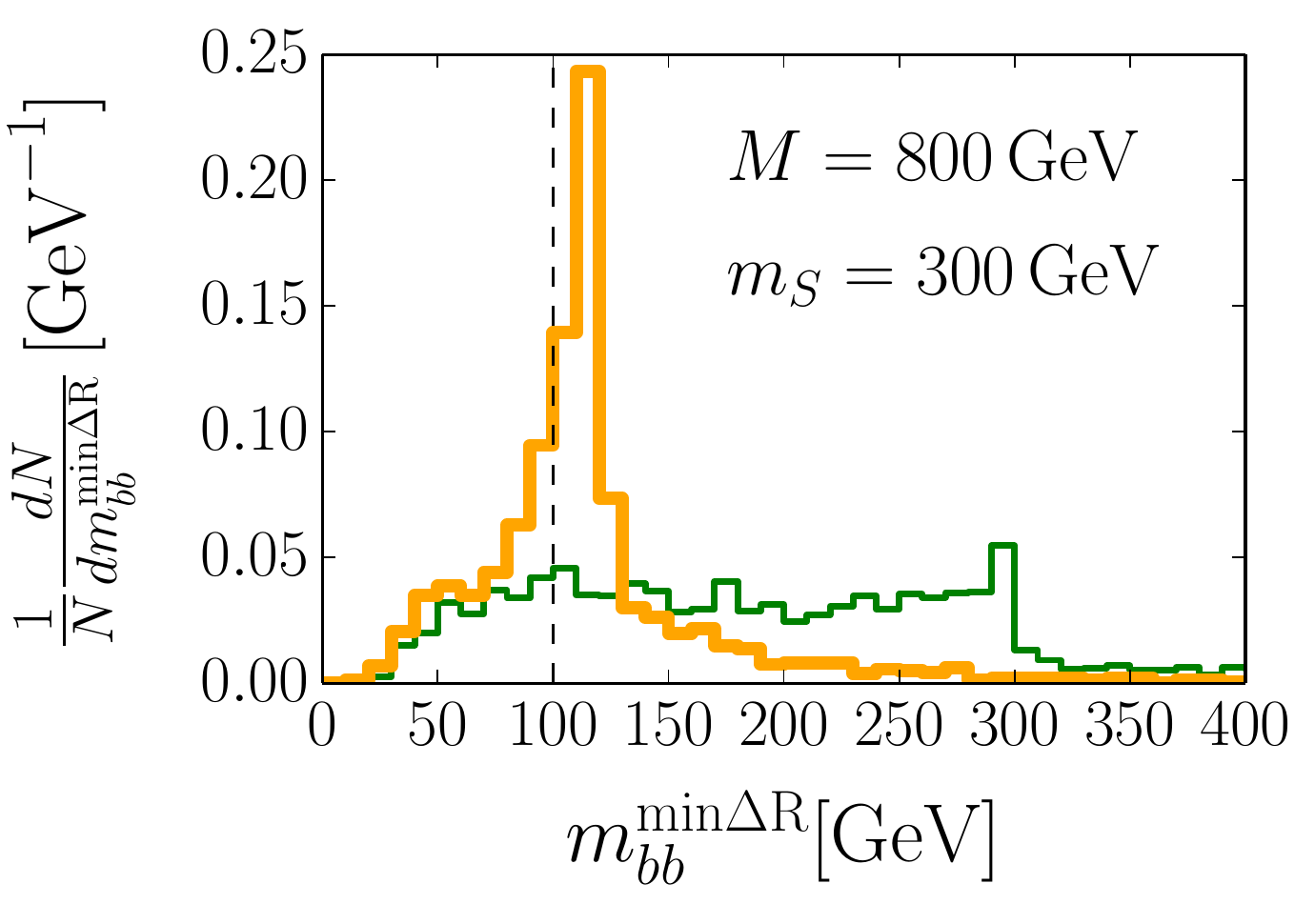}
\end{center}
\caption{\it Left) Normalized distribution of $m_{bb}^{\text{min}\Delta R}$ for $M = 600$ GeV and $m_S = 200$ GeV in the $Ht+Wb$ (thick solid orange) and the $St+Wb$ (thin solid green) channels. The cut on $m_{bb}^{\text{min}\Delta R} > 100$ GeV is depicted with a vertical dashed line. Right) Same as Left) but for $M = 800$ GeV and $m_S = 300$ GeV.}\label{fig:mbb}
\end{figure}
The search for $Ht+X$~\cite{Aad:2015kqa} relies on the Higgs decaying into bottom quarks. Hence, it is also sensitive to $St+X$ events with $S$ decaying into the same final state. In fact, if $m_S$ were of the size of the Higgs mass, the sensitivity of this search for both channels would be similar. A thorough inspection of this search reveals that the only variable that can distinguish both channels is actually $m_{bb}^{\text{min}\Delta R}$, defined as the invariant mass of the two $b$-tagged jets closest in $\Delta R$. It is required to be larger than 100 GeV.

The boost factor of $S$ grows as $\sim M/m_S$. Therefore, the larger its mass, the smaller the fraction of events for which the two bottom quarks resulting from its decay are the closest ones.
Therefore, it could be expected that, for small $M$ and large $m_S$, the sensitivity of this search for $St+X$ reduced with respect to $Ht+X$. However, this effect turns out to be largely compensated by the smaller Higgs decay rate into bottom quarks and the fact that the invariant mass of the closest $b$-tagged jets peaks at much larger values in the $S$ case. This is made explicit in fig.~\ref{fig:mbb} for two different pairs of $M$ and $m_S$. The distributions are obtained out of QCD pair-produced heavy tops generated with \texttt{MadGraph v5}~\cite{Alwall:2014hca} whose SM final state particles are subsequently decayed with  \texttt{Pythia v6}~\cite{Sjostrand:2006za}. The basic cuts of ref.~\cite{Aad:2015kqa}, described in section~\ref{sec:wb}, have been applied. The efficiency with respect to the standard Higgs channel varies from $\sim 1.1$, for $(M, m_S) = (600, 200)$ GeV, to $\sim 1.2$, for $(M, m_S) = (1500, 200)$ GeV.

Similarly, the efficiency of the search for $Zt+X$~\cite{ATLAS-CONF-2017-015} examined in section~\ref{sec:zt} for the $Zt+St$ channel matches also the one for $Zt+Ht$ when $S$ decays to bottom quarks. As a result, we can safely add $\text{BR}(T\rightarrow St)+\text{BR}(T\rightarrow Ht)$ into a single variable that, abusing notation, we will still write as $\text{BR}(T\rightarrow Ht)$. Bounds considering this new channel are then trivial to obtain. Hereafter, we shall only write $St$ explicitly for the case in which $S$ scapes detectors, to be discussed in the next section.

\subsection{$T\rightarrow St, S\rightarrow E_T^\text{miss}$}
\label{sec:et}
The search for $Zt+X$ focuses on the invisible decay of the $Z$ boson. Therefore it is also sensitive to $St+X$ channels whenever $S$ is stable at detector scales. This analysis puts a stringent cut on the amount of $E_T^\text{miss}$. The efficiency depends notably on $M/m_S$, and hence, contrary to the previous case, $\text{BR}(T\rightarrow St)$ can not just be added to $\text{BR}(T\rightarrow Zt)$. Instead, we need to recast the whole analysis. For this matter, we use homemade routines based on a combination of \texttt{MadAnalysis v5}~\cite{Conte:2012fm,Conte:2014zja}, \texttt{ROOT v5}~\cite{Brun:1997pa} and \texttt{Fastjet v3}~\cite{Cacciari:2011ma}. QCD pair-produced heavy tops decaying into $St+Zt, St+Ht$ and $St+Wb$ are generated with \texttt{MadGraph v5}~\cite{Alwall:2014hca} and \texttt{Pythia v6}~\cite{Sjostrand:2006za}. Detector effects are disregarded, although an efficiency of 0.85 (0.9) for selecting electrons (muons) has been simulated.
Among other variables, this analysis relies on $H_{T,\text{sig}}^{\text{miss}}= (H_T^\text{miss}-100~\text{GeV})/\sigma_{H_T^\text{miss}}$, where $H_T^\text{miss}$ stands for the $p_T$ of the vectorial sum of the jet and lepton momenta. Likewise, $\sigma_{H_T^\text{miss}}$ represents the resolution on $H_T^{\text{miss}}$. We take a conservative value of 7 \%. In order to validate the implementation, we recalculate the efficiencies for the $Zt+Zt$ channel. We find very good agreement with those computed in section~\ref{sec:zt}, some of which have been shown in tab.~\ref{tab:efficiencies}.
\begin{figure}[t]
\begin{center}
  \includegraphics[width=0.49\columnwidth]{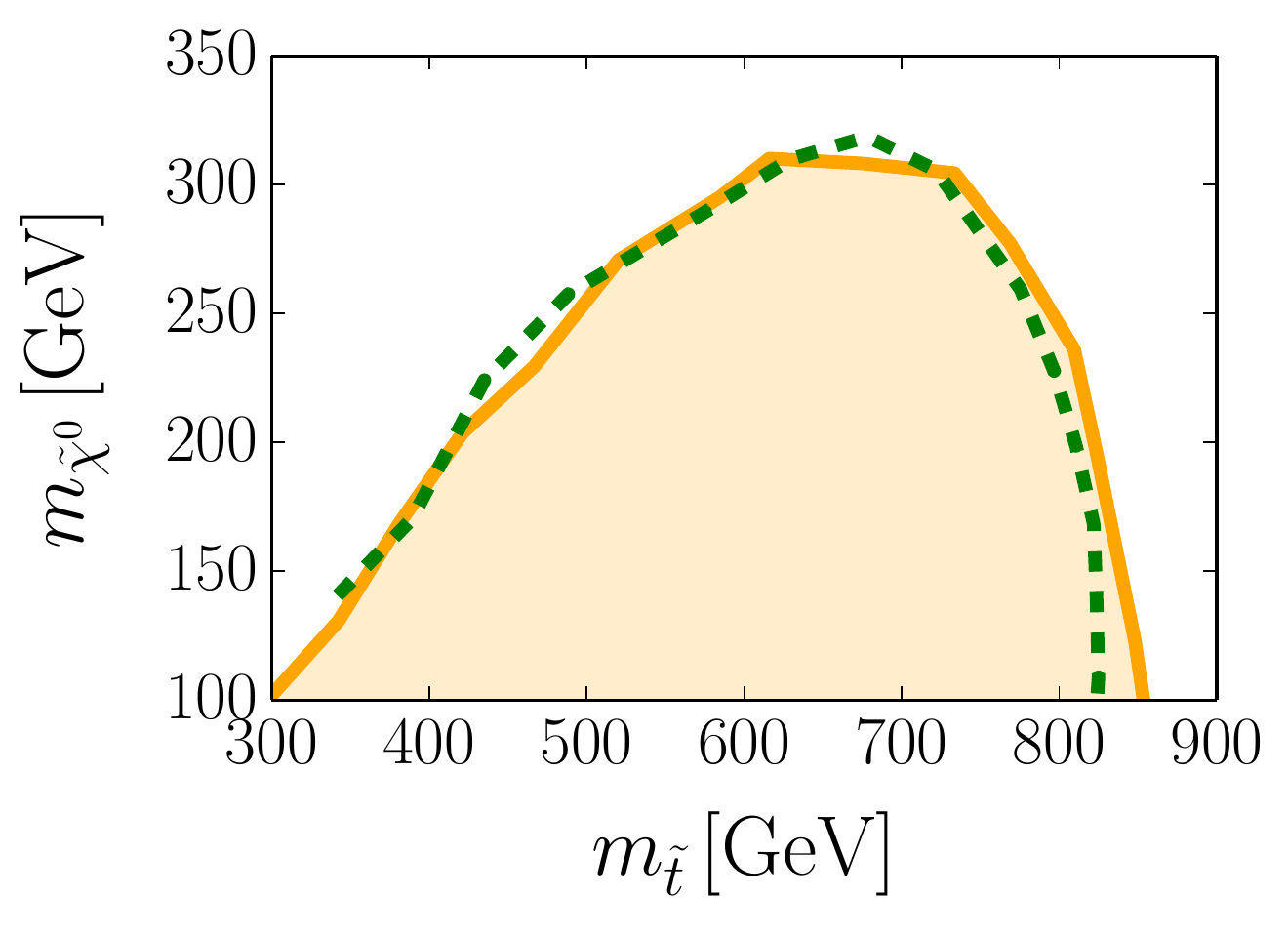}
  \includegraphics[width=0.49\columnwidth]{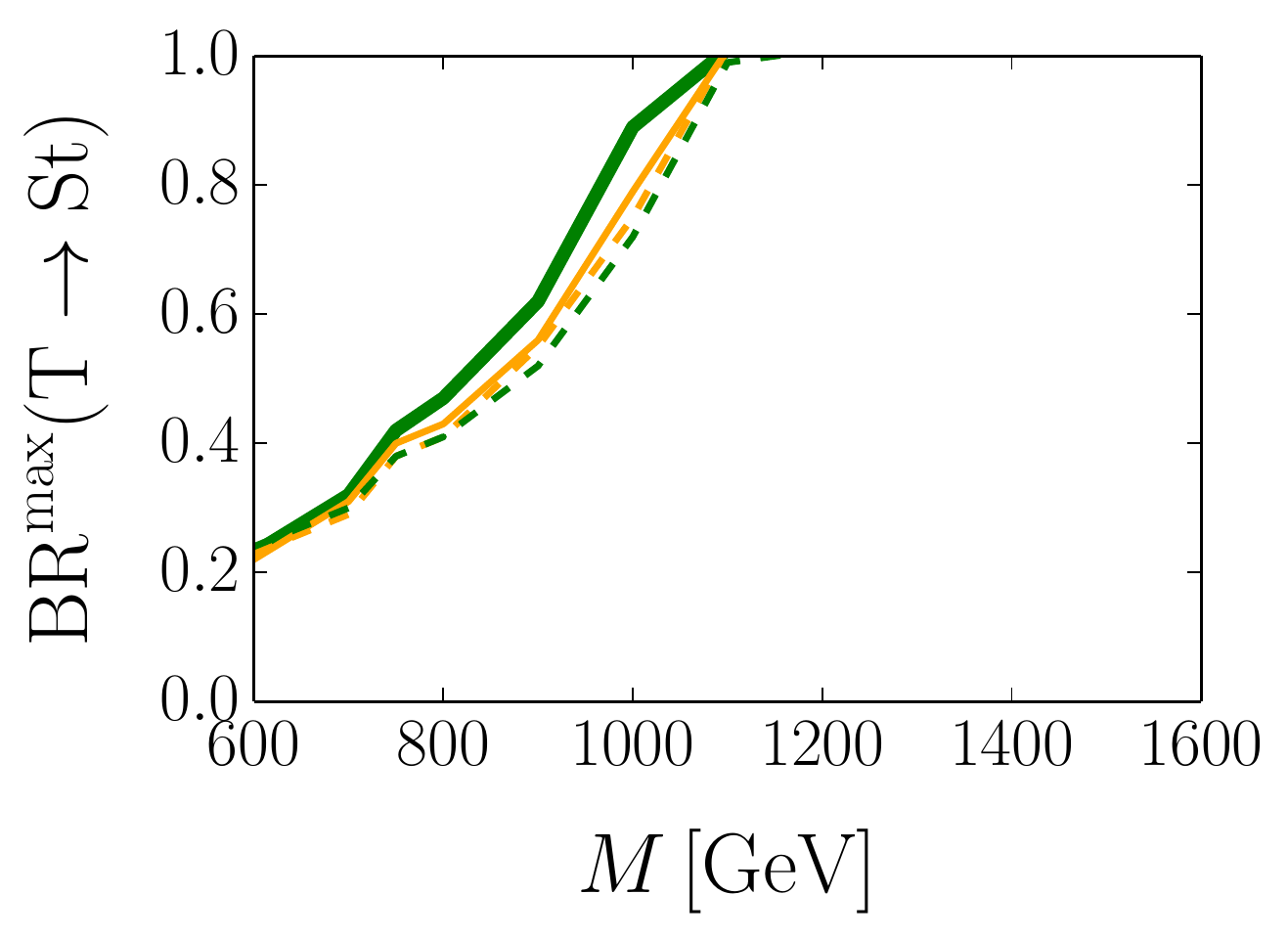}
\end{center}
\caption{\it Left) excluded region in the stop-neutralino mass plane by the CMS search of ref.~\cite{CMS:2016hxa}. The area enclosed by the solid (dashed) orange (green) line is obtained with the recast analysis (by the experimental collaboration). Right) Bound on $\text{BR}(T\rightarrow St)$ for $m_S = 100$ (dashed green), $200$ (dashed orange), $300$ (solid orange) and $400$ (solid green) GeV.}\label{fig:validation}
\end{figure}
Then, we obtain the efficiencies for $St+X$ for different values of $M$ and $m_S$. A representative set of these efficiencies is collected in the different tables of appendix~\ref{app:tables}, together with those computed in the previous sections.

When both top partners decay into $St$, the resulting signal is identical to that explored by searches for stops decaying into tops and neutralinos. We recast one of the latest analyses carried out by CMS~\cite{CMS:2016hxa} at 13 TeV of c.m.e. in the fully-hadronic channel (we also considered the homologous search by the ATLAS Collaboration~\cite{ATLAS:2016jaa}, but the validation was far less convincing). The large top decay rate into jets together with the current performance of boosted techniques make this channel optimal for probing large values of $M$ and $m_S$.

The study relies on 60 statistically independent signal regions. We find that the sensitivity is driven mainly by two of them. Roughly, the first one requires $E_T^\text{miss}$ to be between $250$ and $350$ GeV, as well as $n_t = 0$ and $n_W = 1$. $n_t$ (resp. $n_W$) is the number of fat jets with mass within 110  and 120 (resp. 60 and 110) GeV and $p_T > $ 400 (resp. 200) GeV. Fat jets are clustered out of hadronic traces with the anti-$k_t$ algorithm~\cite{Cacciari:2008gp} with $R = 0.8$ and subsequently re-clustered with the Cambridge-Aachen algorithm~\cite{Dokshitzer:1997in} with the same $R$. The analysis reports the observation of 14 events, while only 9.6 are expected from the SM alone (see tab.~\ref{tab:effs}). Consequently, any model giving more than 13 events in this signal region is excluded at the 95 \% C.L. The second signal region, instead, requires $E_T^\text{miss}>500$ GeV, $n_t \geq 1$ and $n_W \geq 1$. The number of observed and expected events are 1 and 0.16, respectively. This sets an upper limit on the number of allowed signal events after cuts of 5. In both cases, at least five $R=0.4$ anti-$k_t$ jets and at least two $b$-tagged jets have to be present.

\begin{table}[t]
 \begin{center}
\begin{adjustbox}{width=0.95\textwidth}
\footnotesize
\begin{tabular}{||c|cccc||}\hline
 & $Ht + X$ & $Wb + X$ & $Zt + X$ & CMS \\  
 \hline
 \textit{c.m.e.} [TeV] & 8 & 8 & 13 & 13 \\[0.1cm]\hline
  & $HtHt$, $HtWb$,& $WbWb$ & $ZtZt$, $ZtHt$, & $StSt$ \\
Channels & $HtZt$ &  & $ZtWb$, $StHt$,  &  \\
  &  &  & $StZt$, $StWb$ & \\\hline
  $N_b$ & 81 & $27.6$ & $6.5$ & 9.6, 0.16 \\
  $N_d$ & 84 & 30 & $7$ & 14, 1 \\
  $N_s$ & 22 & 15 & $8$ & 13, 5\\
 \hline
 \end{tabular}
 \end{adjustbox}
 \end{center}
 \caption{\it Experimental analyses considered in this work and channels for which efficiencies have been estimated. The number of observed and expected SM events in each signal region are also shown. Any signal leading to more than $N_s$ events in at least one analysis is excluded, although stronger bounds are obtained if the different analyses are combined into a shape analysis. See the text for details.}\label{tab:effs}
\end{table}

\begin{figure}[ht]
\begin{center}
  \includegraphics[width=0.49\columnwidth]{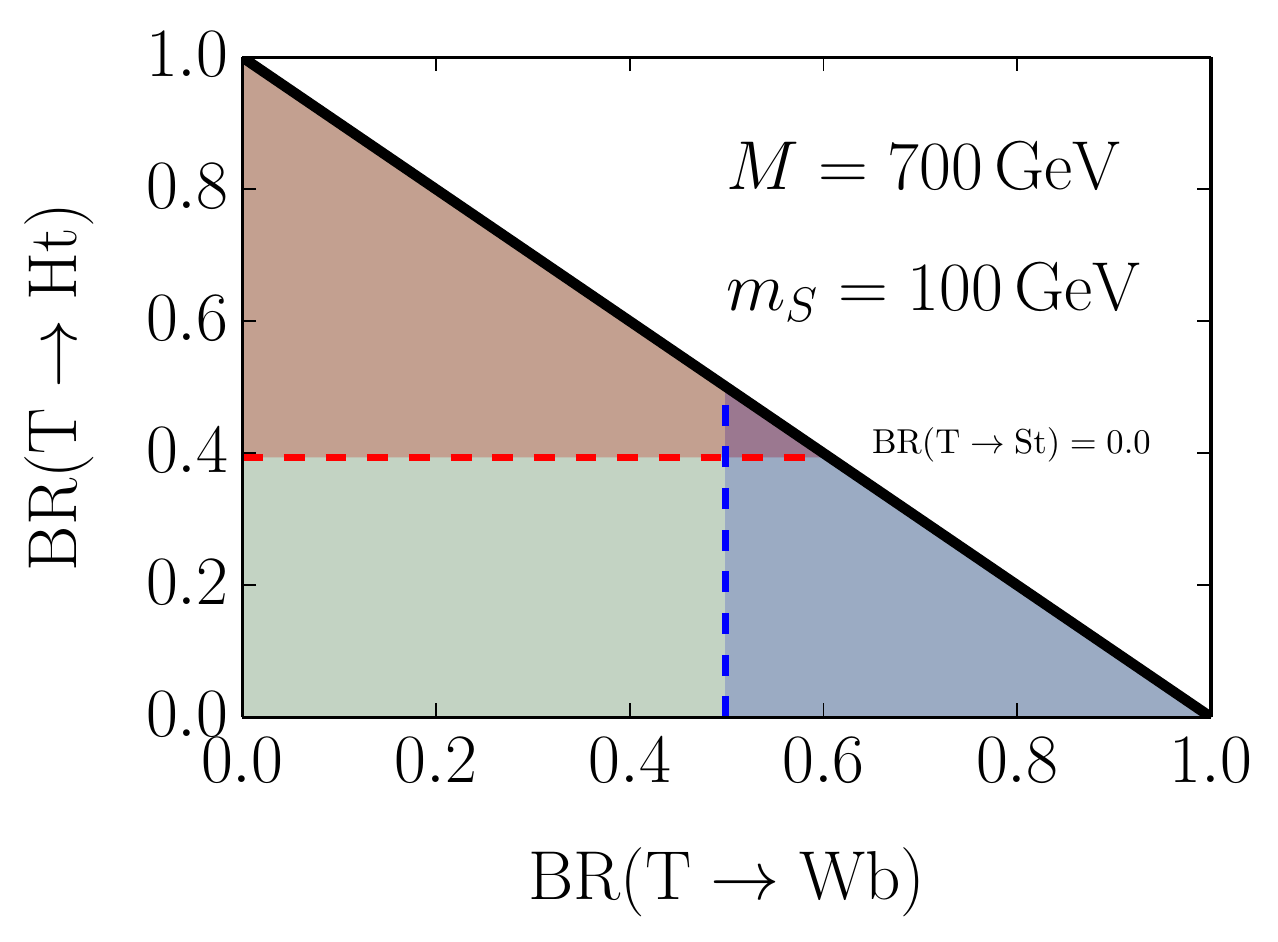}
  \includegraphics[width=0.49\columnwidth]{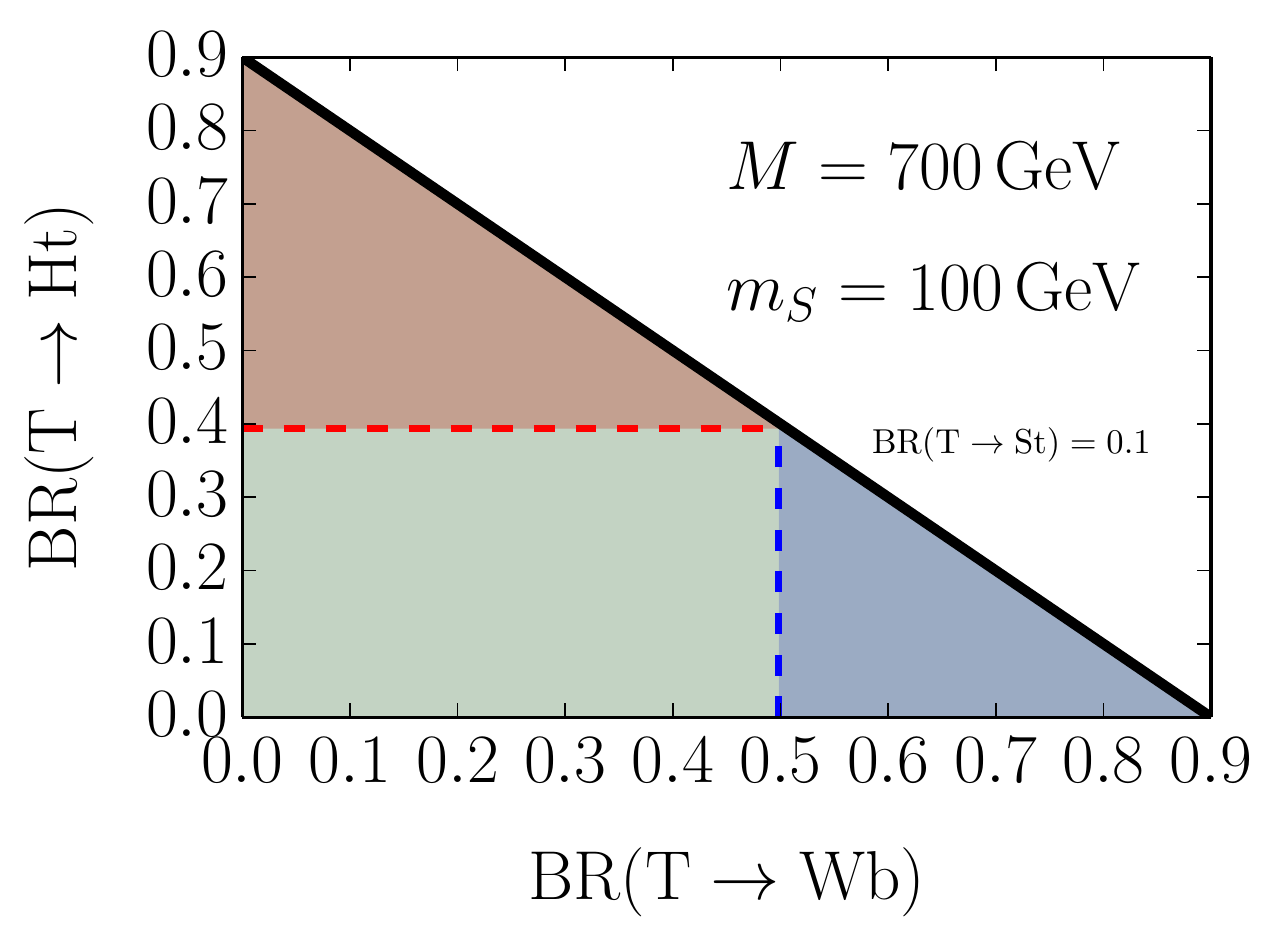}
  \includegraphics[width=0.49\columnwidth]{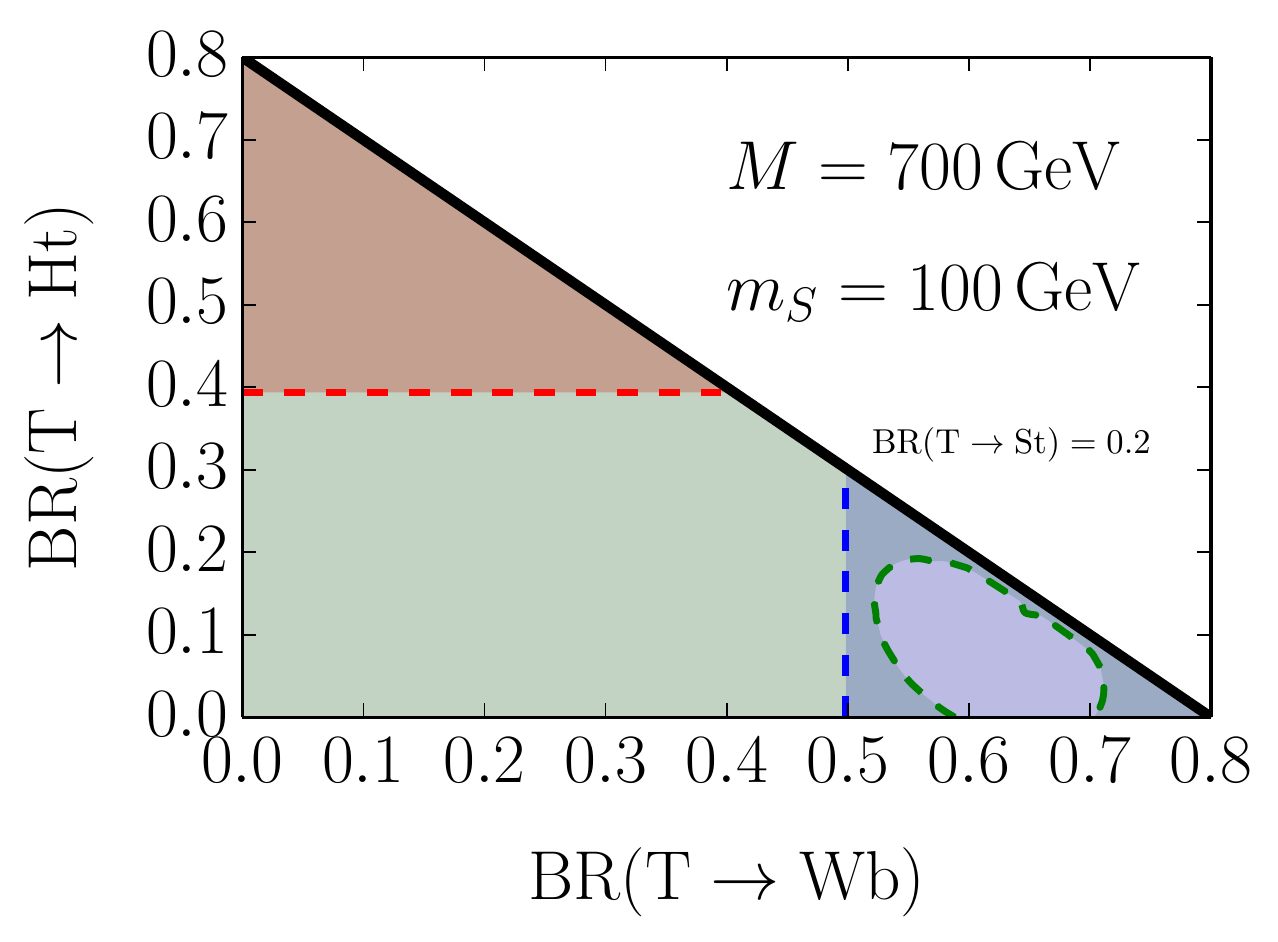}
  \includegraphics[width=0.49\columnwidth]{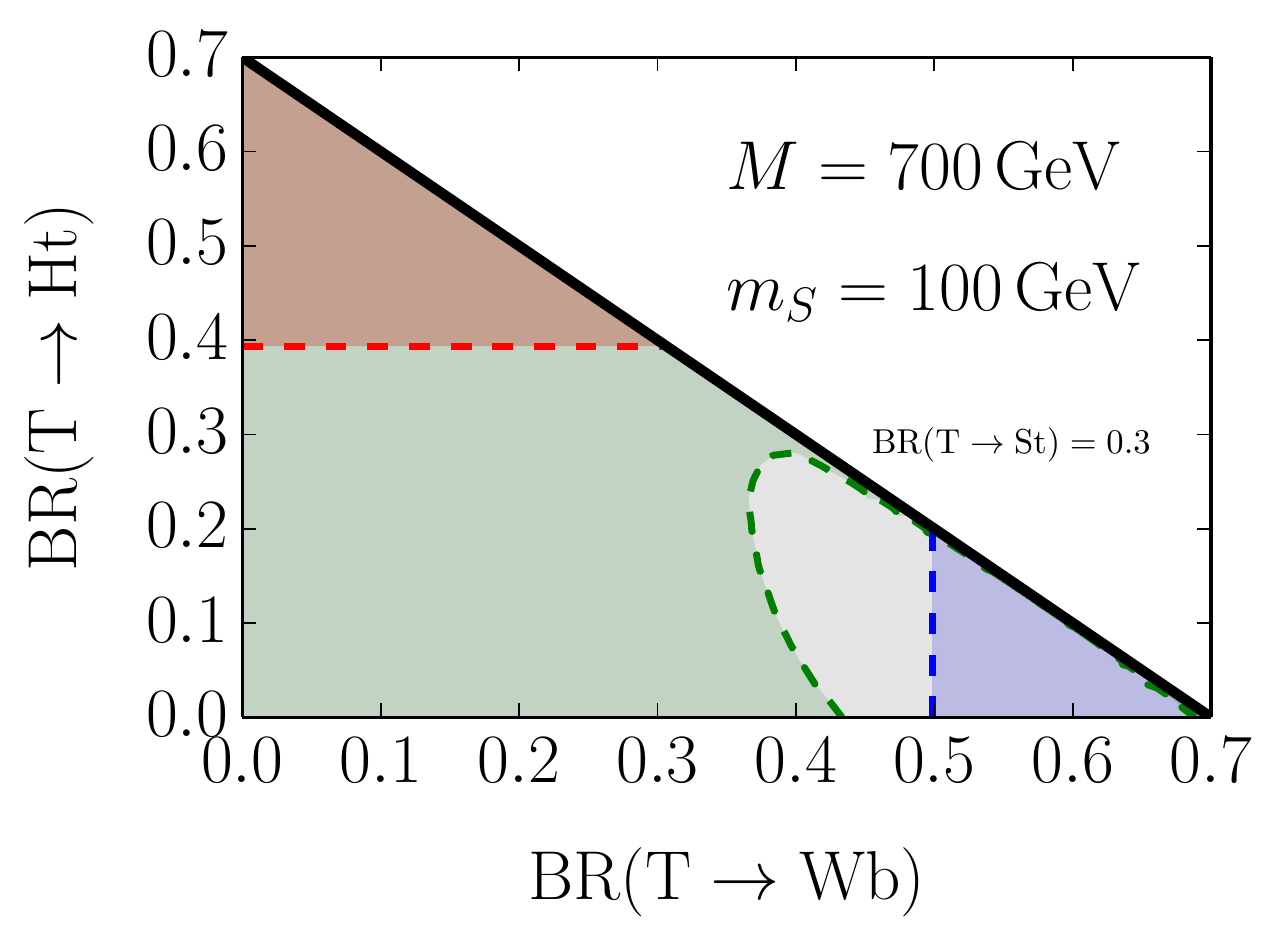}
\end{center}
\caption{\it Excluded regions in the space of branching ratios for $M = 700$ GeV and $m_S = 100$ GeV. The gray area results from a combined statistical analysis of all signal regions of tab.~\ref{tab:effs}. In the green area, the $St+X$ events are disregarded. The red region is excluded by the analysis of ref.~\cite{ATLAS-CONF-2016-104}. Likewise, the blue region is excluded by the analysis of ref.~\cite{ATLAS-CONF-2016-102}.}\label{fig:700}
\end{figure}

For the validation we produce again pairs of stops with \texttt{MadGraph v5} that are subsequently decayed \textit{only} into tops and neutralinos with \texttt{Pythia v6}.
We then obtain the number of survivor signal events in each of the 60 signal regions considered in the experimental analysis. We combine all them into a single statistics which is compared with the provided number of expected and observed events via a CL$_s$ analysis. The excluded region in the plane of stop and neutralino masses ($m_{\tilde{t}}-m_{\tilde{\chi}^0}$) is depicted by the orange region in the left panel of fig.~\ref{fig:validation}. The limits reported by the experimental collaboration are also shown for comparison, the agreement being apparent.
We can therefore obtain the efficiency of this analysis for $St+St$ events for different values of $m_S$. As the rest of efficiencies, these are reported in appendix~\ref{app:tables}.
In the right panel of fig~\ref{fig:validation} we plot the maximum value of $\text{BR}(T\rightarrow St)$ that is allowed by this analysis when it is applied to the $St+St$ channel alone taken into account only the two main important signal regions. Note that values of $M$ as large as 1100 GeV are excluded for $\text{BR}(T\rightarrow St) = 1$. Much lighter resonances are still allowed if the branching ratio is smaller. In that case, however, decays into SM particles are also present. Consequently both standard and exotic $T$ decays have to be considered at once. We address this point in the next section.

 \section{Final results}\label{sec:final}
\begin{figure}[t]
\begin{center}
  \includegraphics[width=0.49\columnwidth]{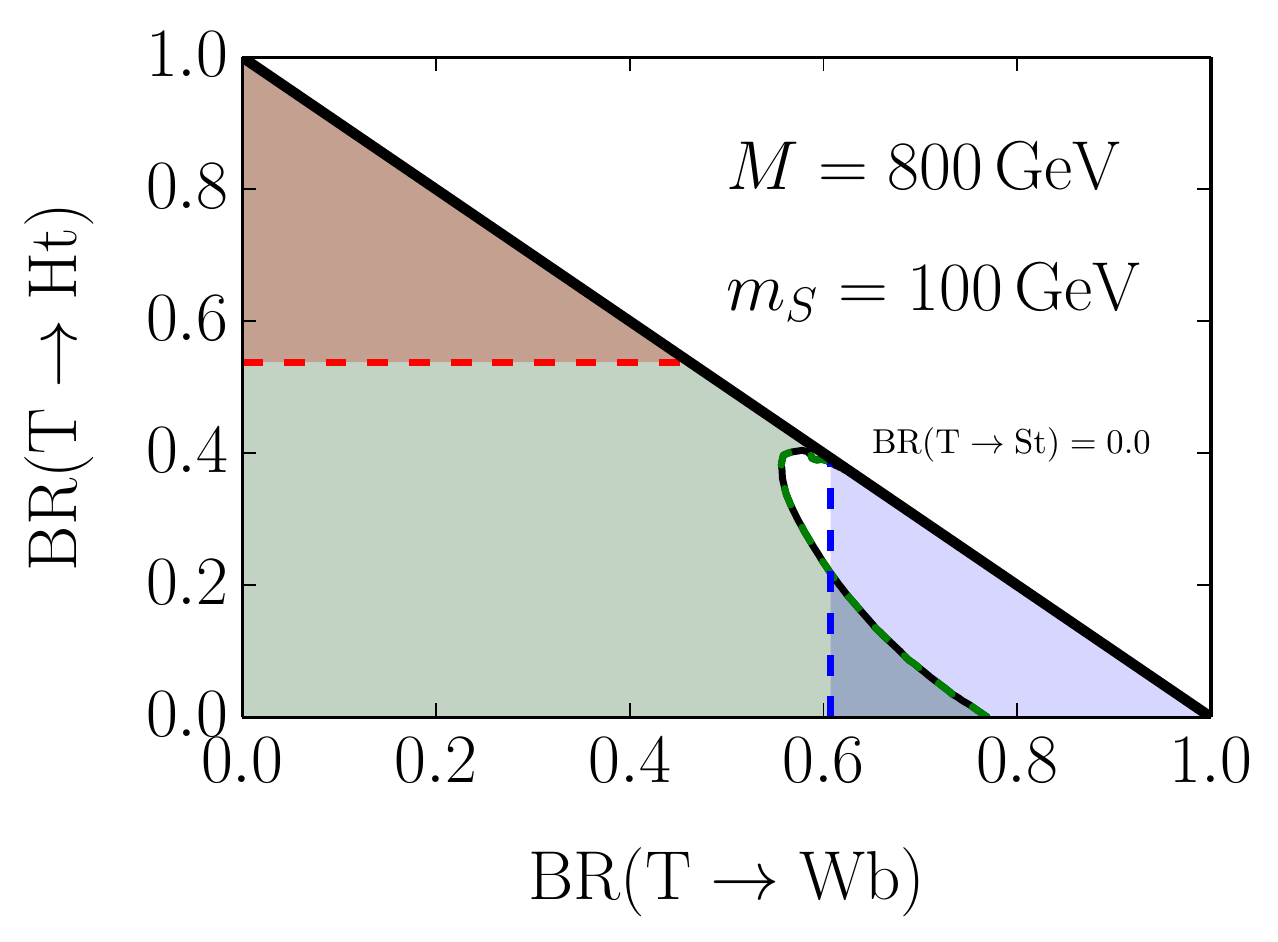}
  \includegraphics[width=0.49\columnwidth]{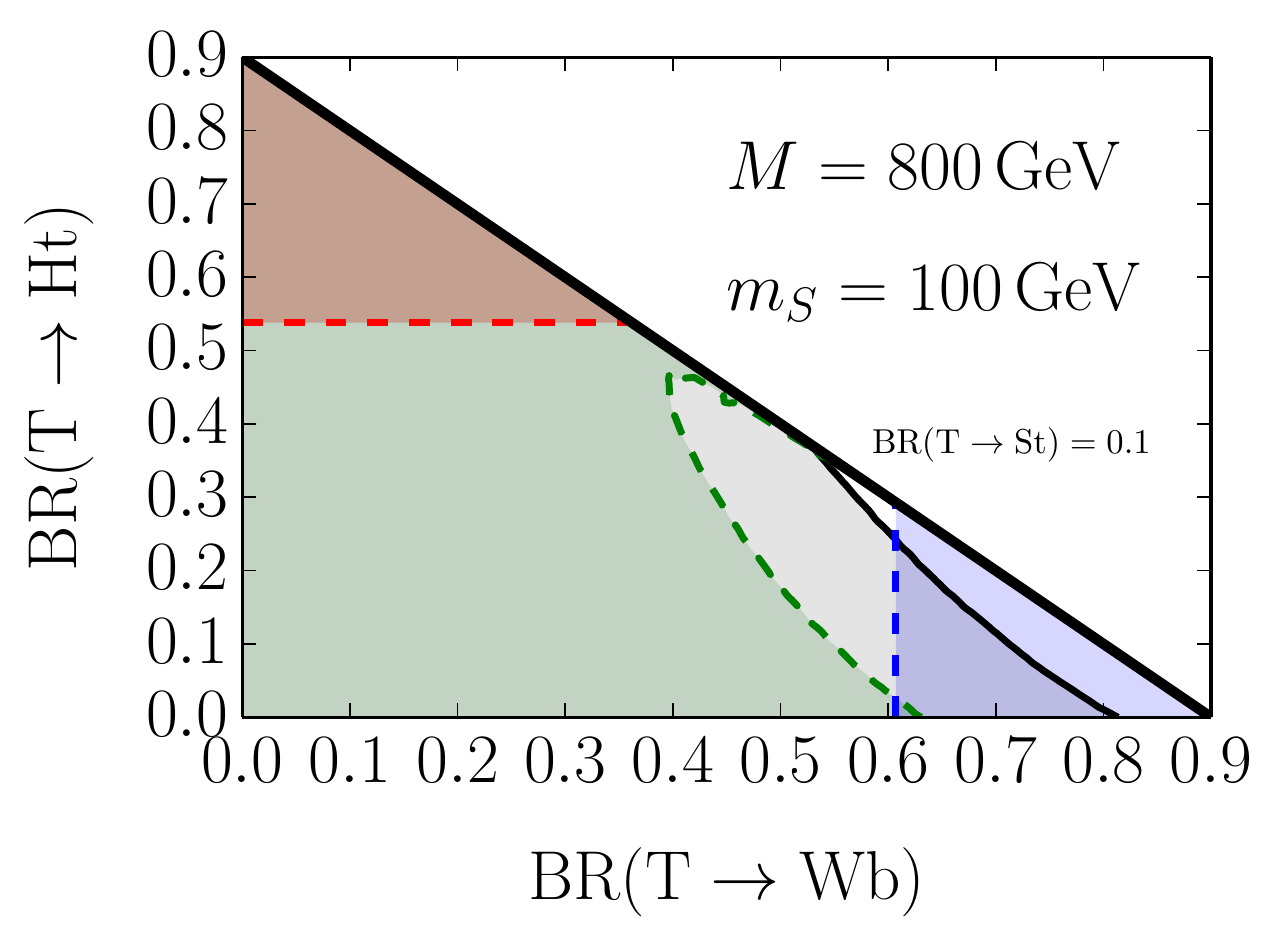}
  \includegraphics[width=0.49\columnwidth]{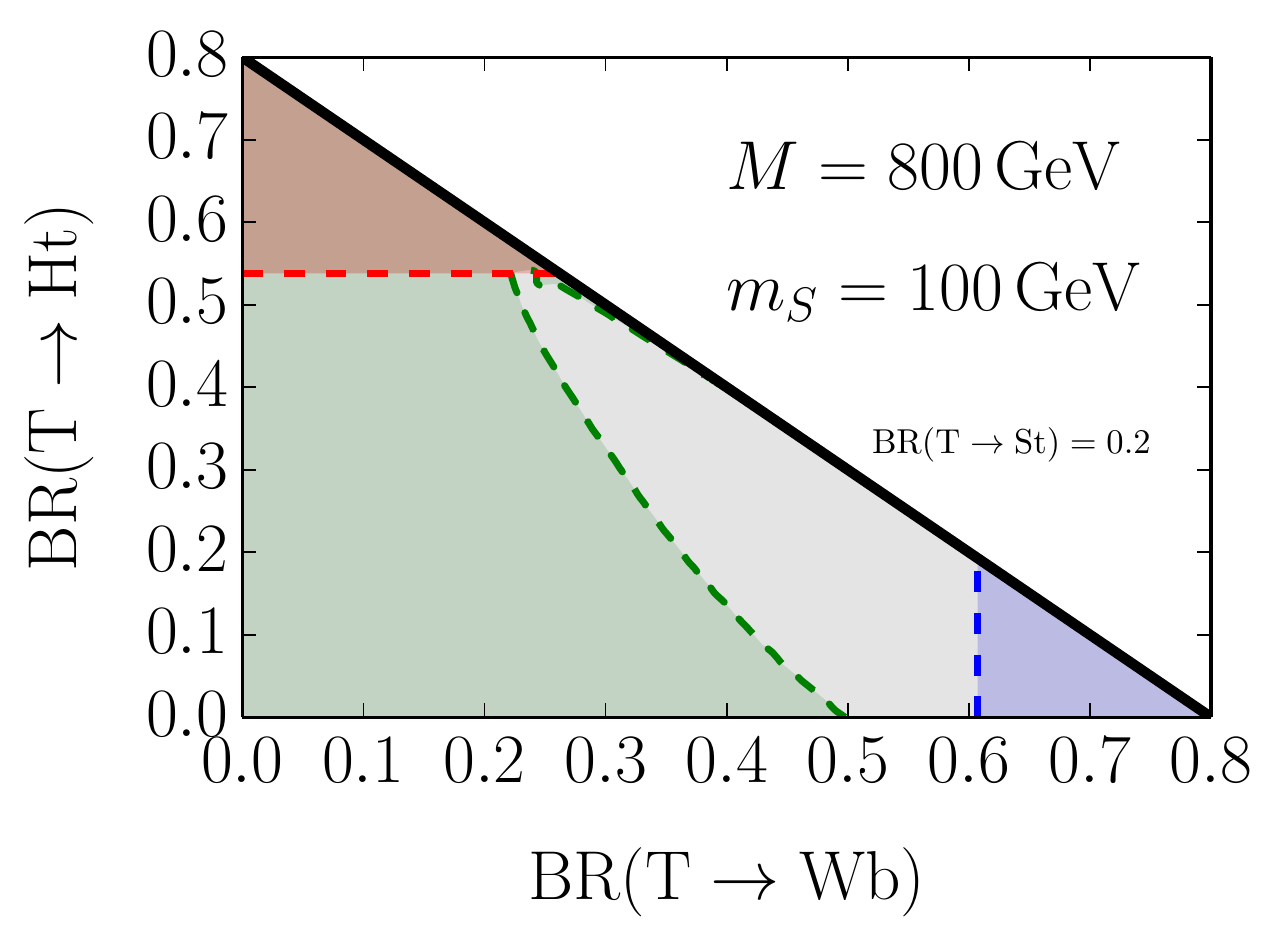}
  \includegraphics[width=0.49\columnwidth]{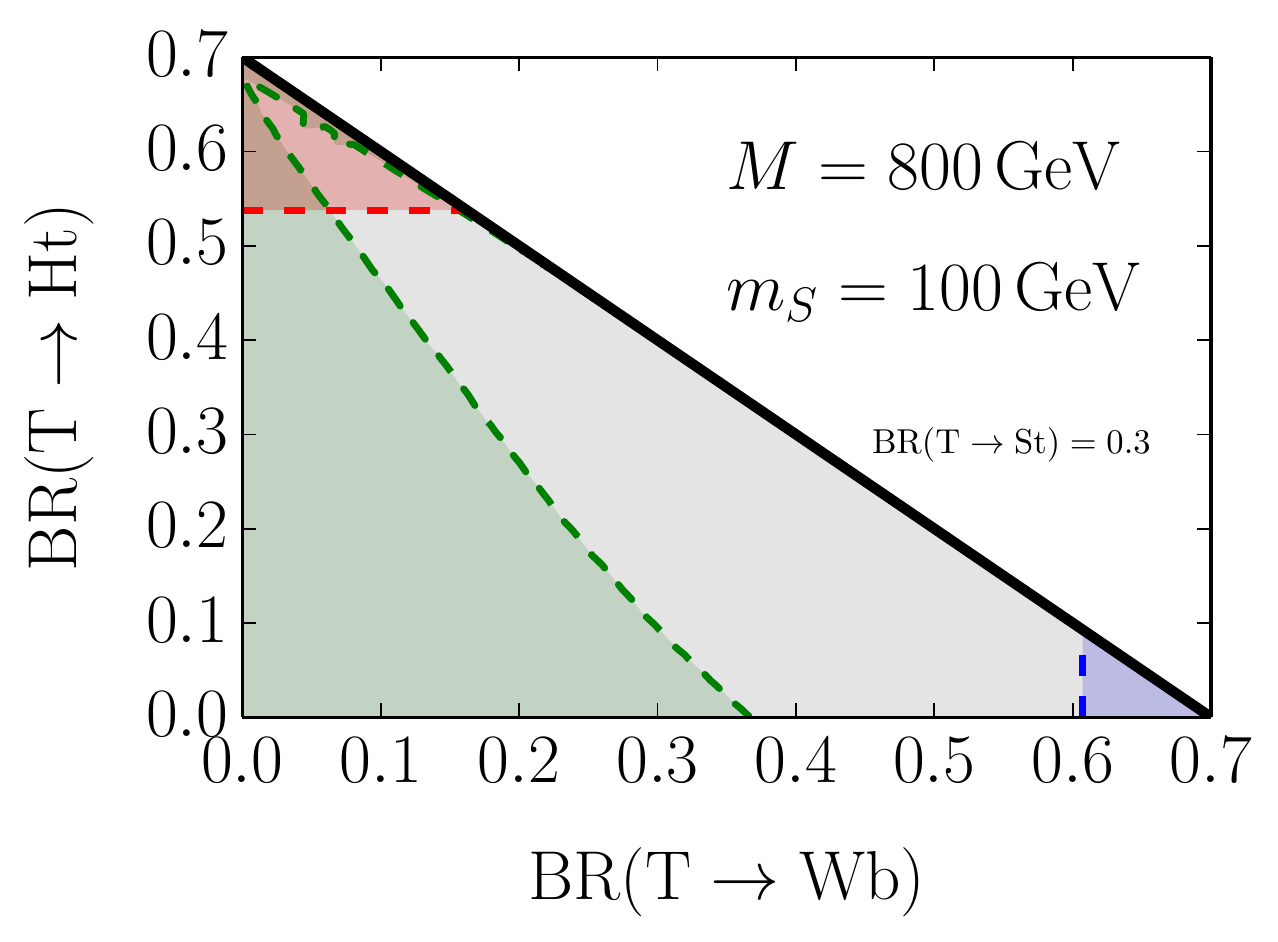}
\end{center}
\caption{\it Same as fig.~\ref{fig:700} but for $M = 800$ GeV.}\label{fig:800}
\end{figure}

\begin{figure}[t]
\begin{center}
  \includegraphics[width=0.49\columnwidth]{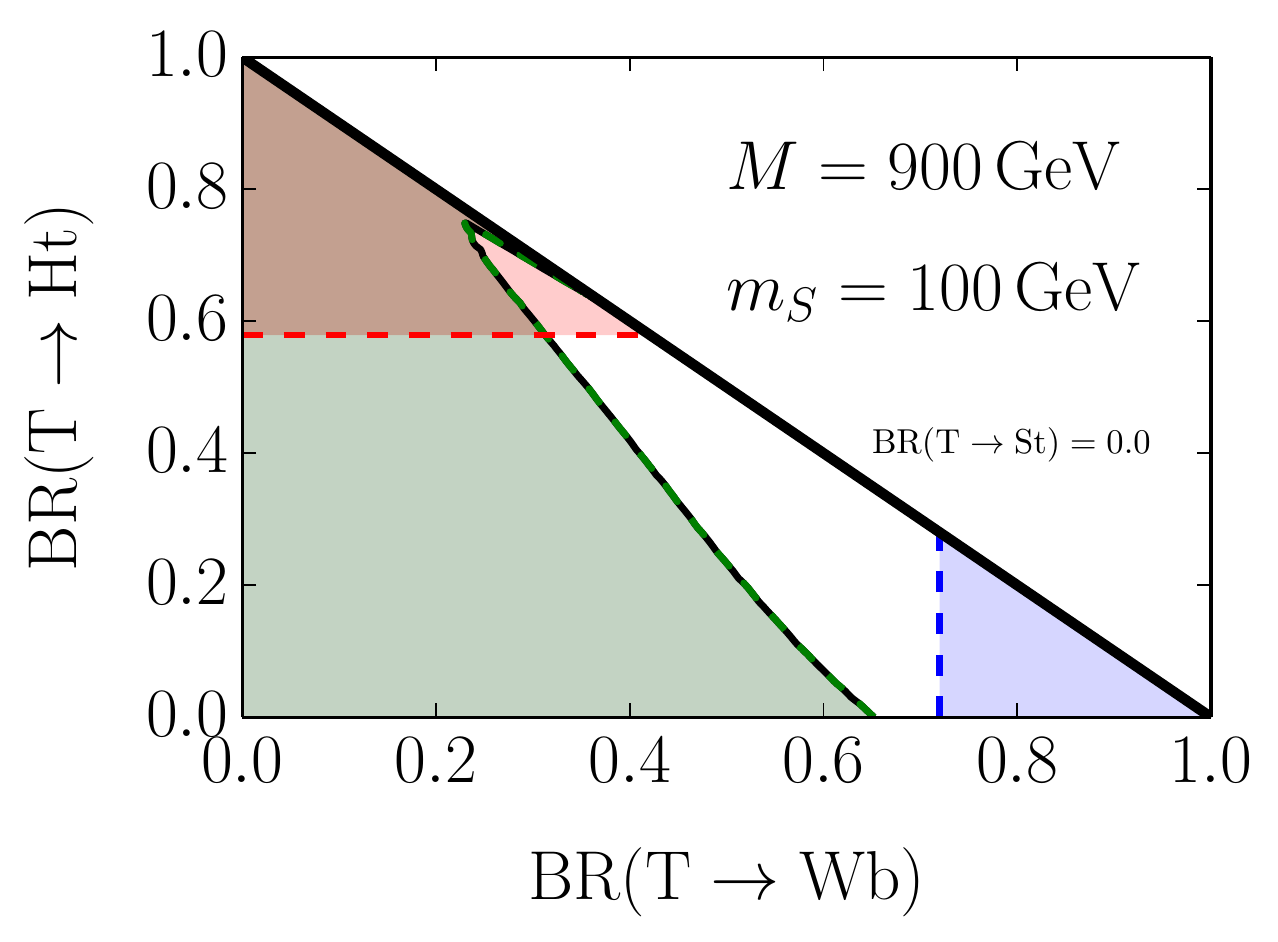}
  \includegraphics[width=0.49\columnwidth]{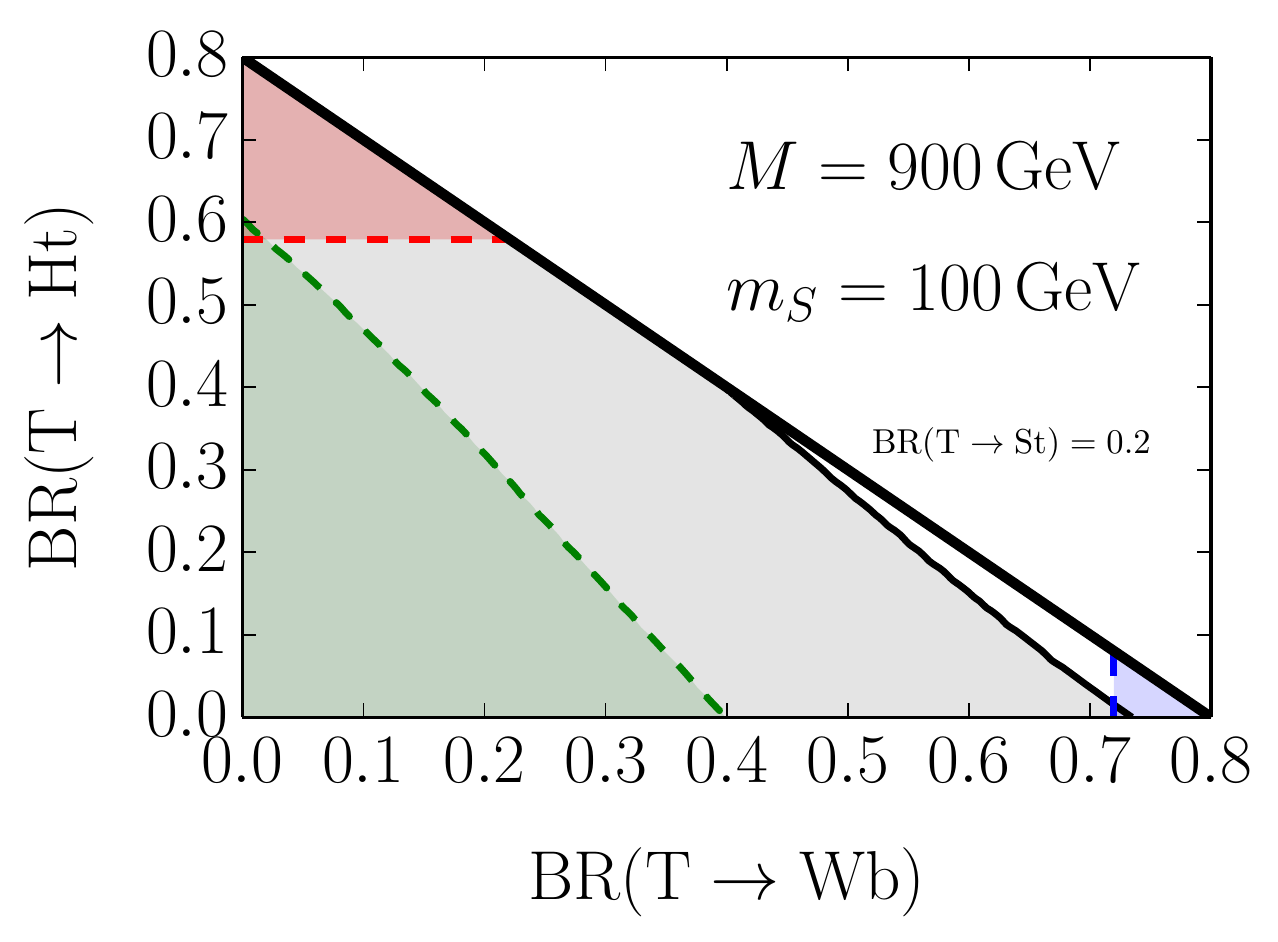}
  \includegraphics[width=0.49\columnwidth]{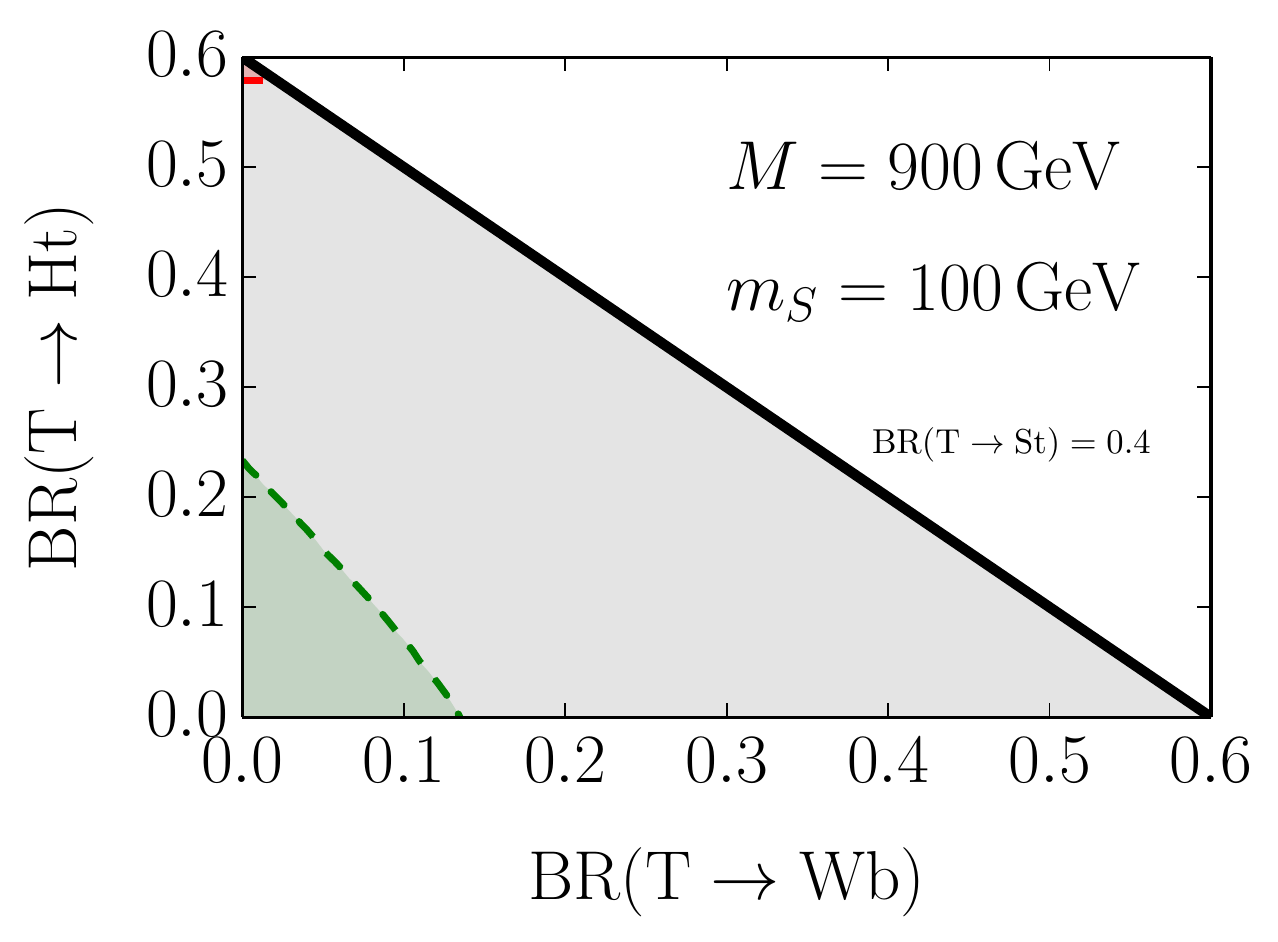}
  \includegraphics[width=0.49\columnwidth]{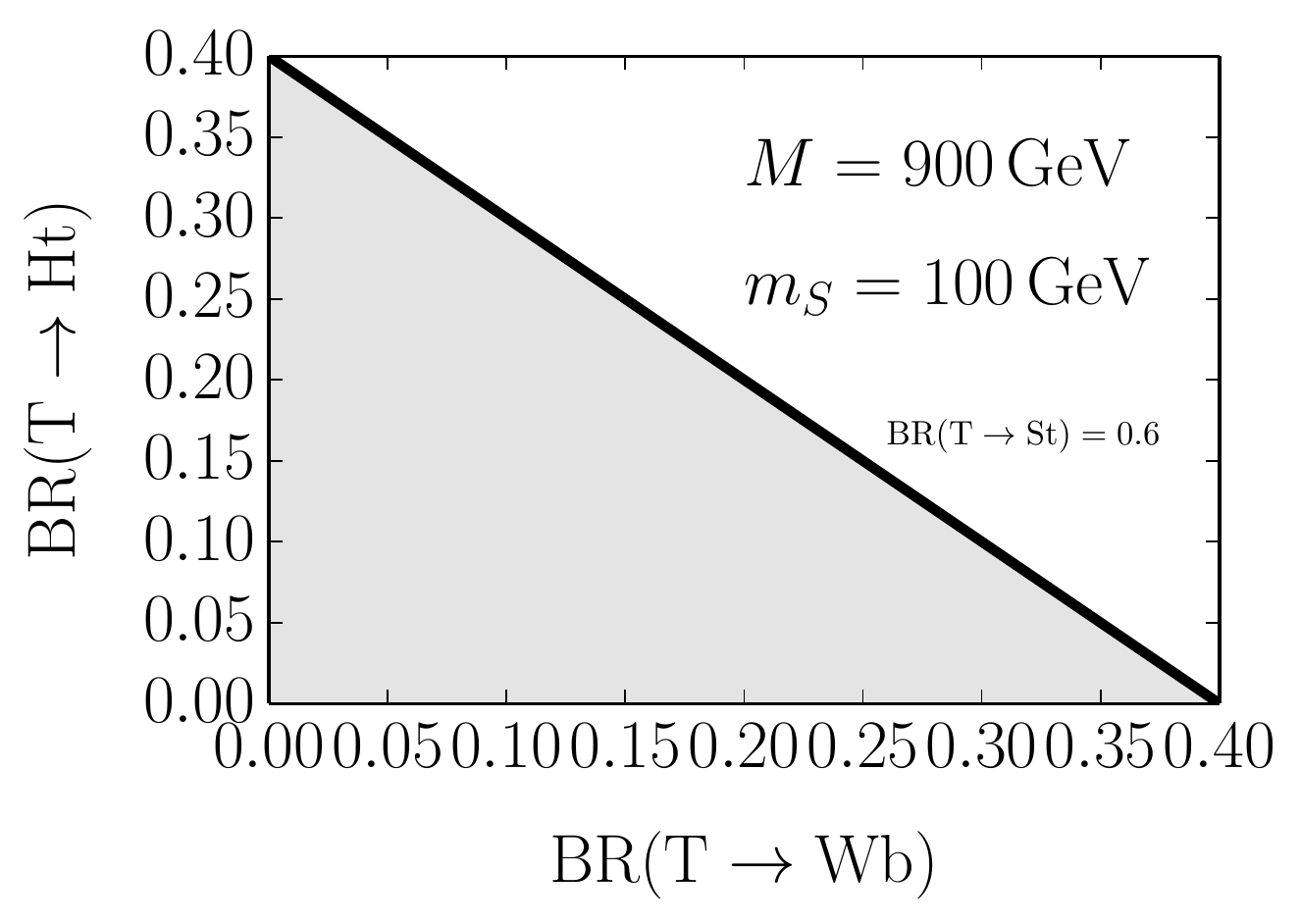}
\end{center}
\caption{\it Same as fig.~\ref{fig:700} but for $M = 900$ GeV.}\label{fig:900}
\end{figure}

\begin{figure}[t]
\begin{center}
  \includegraphics[width=0.49\columnwidth]{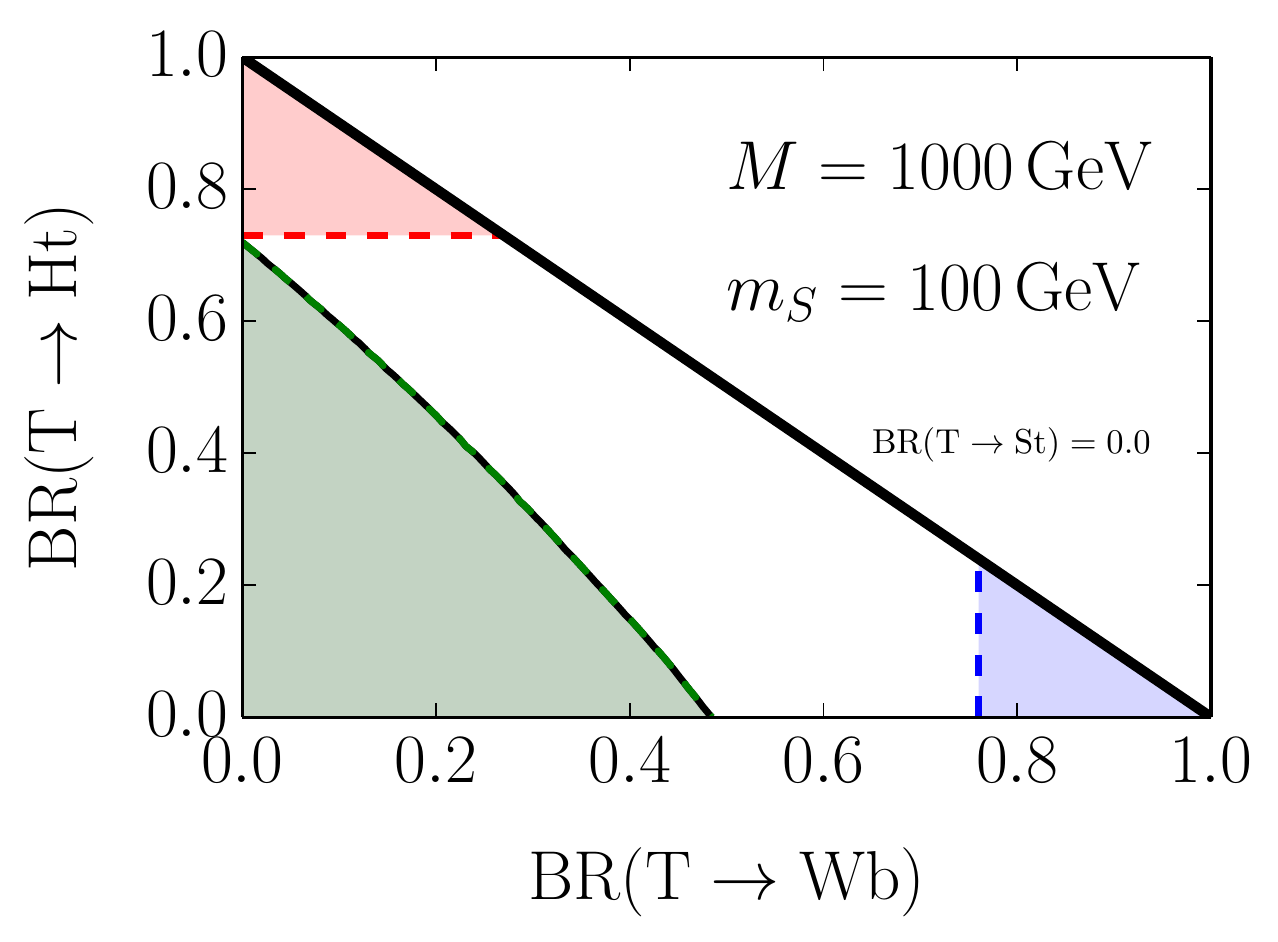}
  \includegraphics[width=0.49\columnwidth]{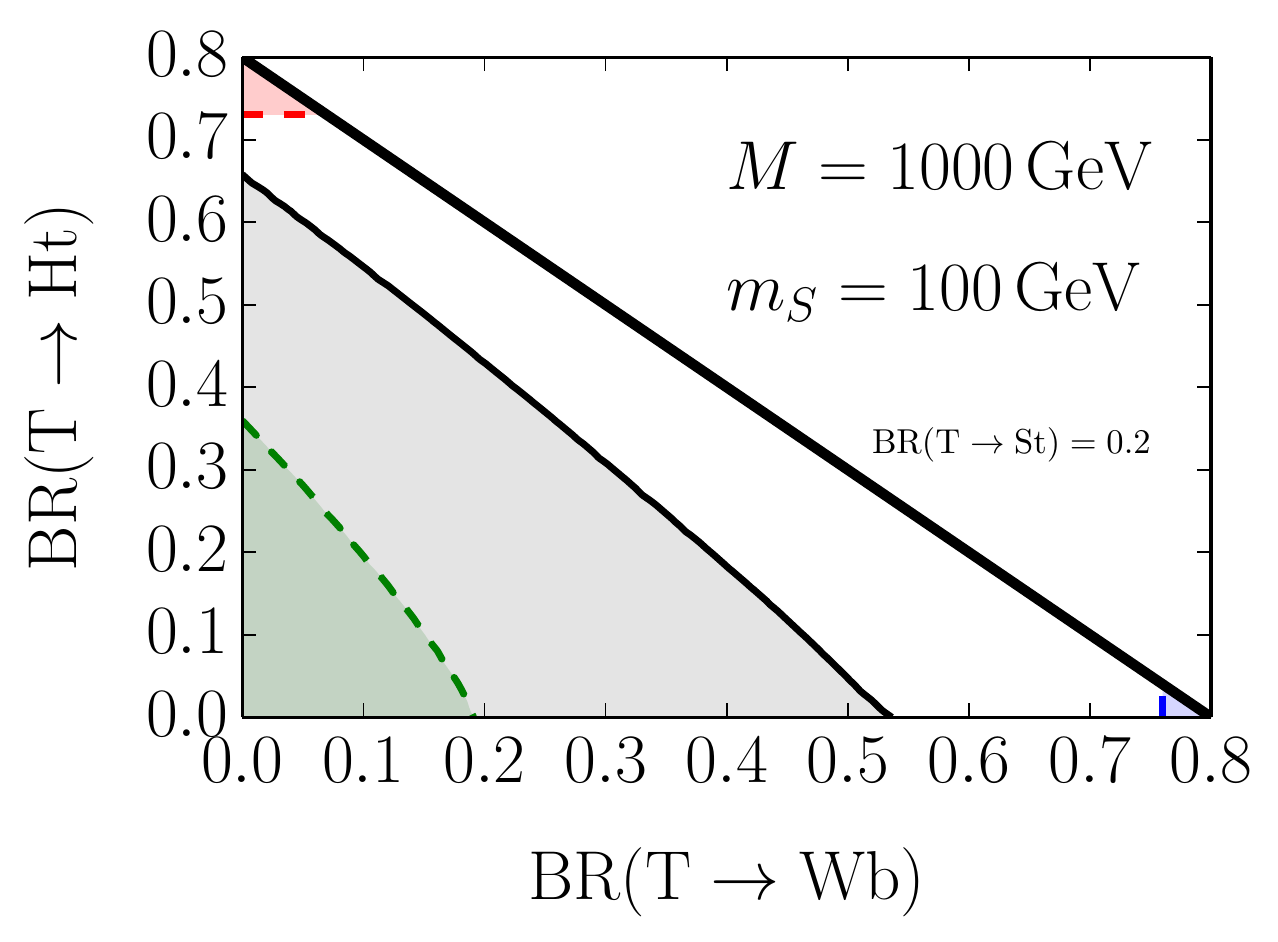}
  \includegraphics[width=0.49\columnwidth]{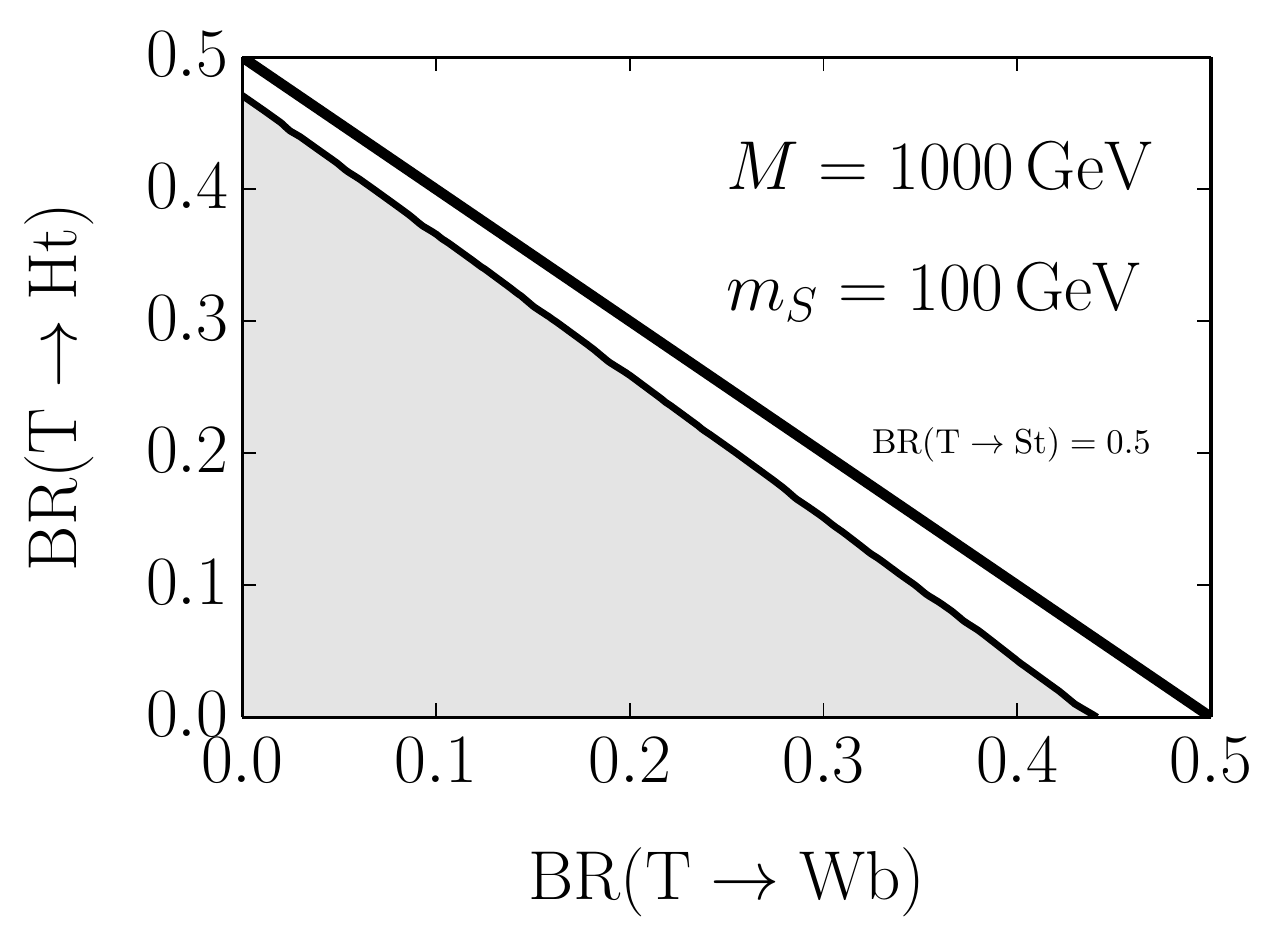}
  \includegraphics[width=0.49\columnwidth]{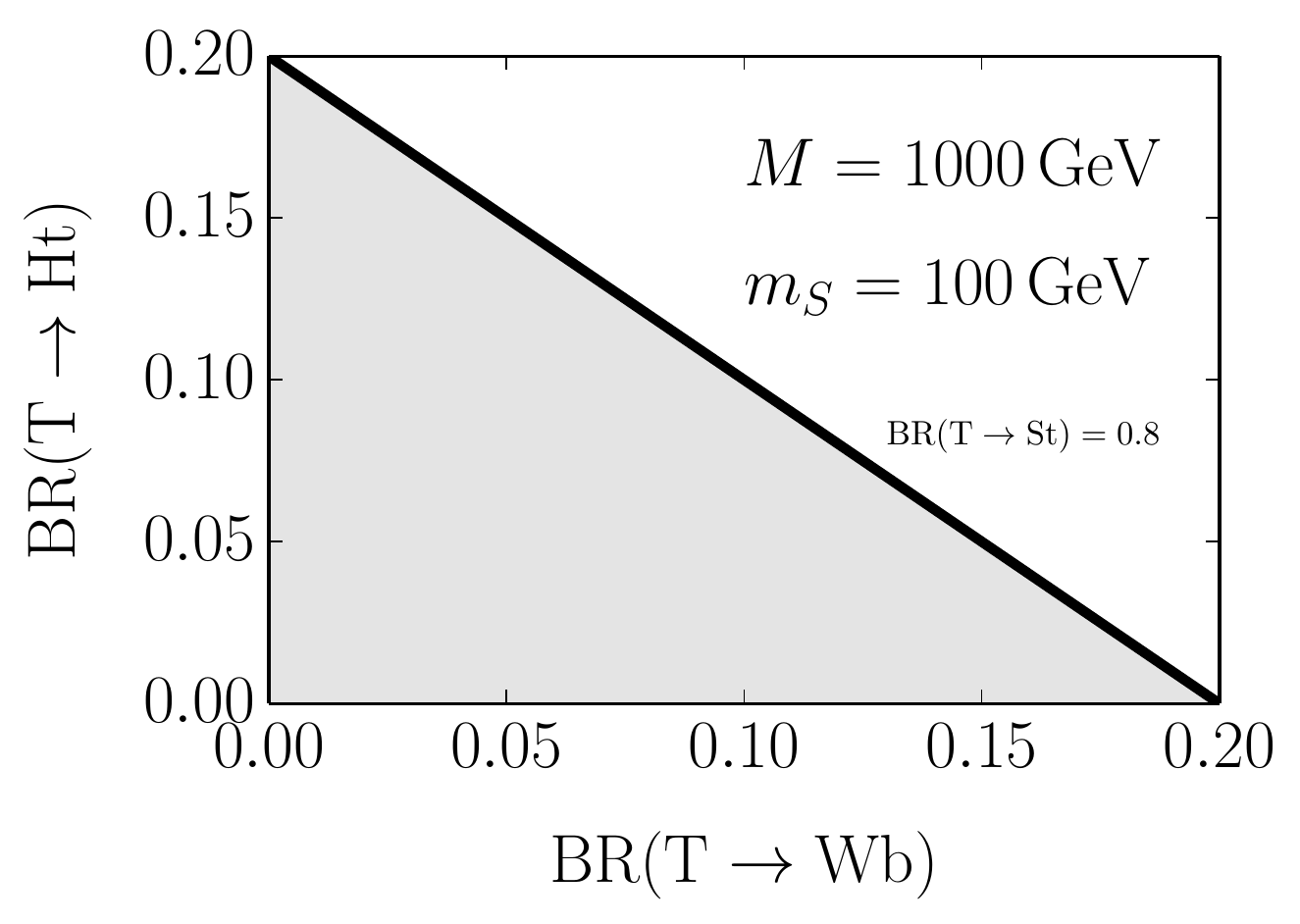}
\end{center}
\caption{\it Same as fig.~\ref{fig:700} but for $M = 1000$ GeV.}\label{fig:1000}
\end{figure}

At this stage, the number of signal events passing the cuts of the analyses reported in tab.~\ref{tab:effs} can be computed for arbitrary branching ratios of a heavy top $T$ into $Zt, Ht, Wb$ ot $St$. We recall that $St$ stands for an invisible $S$, while the $\text{BR}(T\rightarrow St, S\rightarrow b\overline{b})$ is included in $\text{BR}(T\rightarrow Ht)$. The number of expected and observed events for each analysis are also written in the table. Given this information, bounds combining all these searches can be obtained by using the CL$_s$ method as described in section~\ref{sec:zt}. These are shown in figs.~\ref{fig:700},~\ref{fig:800},~\ref{fig:900} and ~\ref{fig:1000} for different values of $M$ and different choices of $\text{BR}(T\rightarrow St)$. Given the small dependence of searches for $St+St$ on the value of $m_S$ (see the right panel of fig.~\ref{fig:validation}), we have fixed $m_S=100$ GeV in all plots. We have also assumed
\begin{equation}
\text{BR}(T\rightarrow Ht) +  \text{BR}(T\rightarrow Wb) + \text{BR}(T\rightarrow Zt) + \text{BR}(T\rightarrow St) = 1~.
\end{equation}
The different colors have the following meanings. The \textit{gray} area enclosed by the solid black line shows the region excluded after combining all the (statistically independent) signal regions presented in tab.~\ref{tab:effs}. In the \textit{green} one, enclosed by the dashed green line, we do the same, but neglecting the events resulting from $T\rightarrow St$. This region gives an idea of how constraining current searches are if $T$ decays also into some elusive channel (for example into very soft final state particles). In other words, it reflects departures from eq.~\ref{eq:brs} by an amount of $\text{BR}(T\rightarrow St)$. Clearly, in the limit in which $\text{BR}(T\rightarrow St) = 0$, this region coincides with the gray one. In the limit $\text{BR}(T\rightarrow St) = 1$, this region is empty. The \textit{red} region enclosed by the dashed red line represents the area excluded by searches for $Ht+X$ at 13 TeV of c.m.e., as explained in section~\ref{sec:ht}. Finally, the blue region enclosed by the dashed blue line represents the area excluded by searches for $Wb+X$ at 13 TeV of c.m.e., as explained in section~\ref{sec:wb}. Let us discuss these results case by case.

\textbf{$\mathbf{M = 700}$ GeV:} All combinations of branching ratios are excluded, independently on the value of $\text{BR}(T\rightarrow St)$. Note that we do not show results for values of this rate larger than 0.3. The reason is that, according to the plot in the right panel of fig.~\ref{fig:validation}, such scenarios are already excluded by searches for stops decaying into neutralinos.

\textbf{$\mathbf{M = 800}$ GeV:} In this case, a small region of the plane when $\text{BR}(T\rightarrow St)< 0.1$ is still allowed by the data we are considering. We stress, however, that this region disappears when all signal regions in the search for $Ht+X$ (see ref.~\cite{ATLAS-CONF-2016-104}) are considered. It is also worth to emphasize the power of combining, in a single statistical analysis, the data from heavy tops with standard decays with those for top partners with exotic decays. Indeed, let us focus on the case $\text{BR}(T\rightarrow St) = 0.2$ (left lower panel). The region excluded by searches for top partners decaying into $Zt, Ht$ or $Wb$ is depicted by the green area. Clearly, this is far from being the whole parameter space. On the other hand, $\text{BR}(T\rightarrow St) = 0.2$ is perfectly allowed by searches for $St+St$ (right panel of fig.~\ref{fig:validation}). The combination of the various channels, however, gives much more stringent constraints than the mere superposition of those coming from the different searches alone.

\textbf{$\mathbf{M = 900}$ and 1000 GeV:} In both cases, there are sizable regions of the parameter space not excluded by current searches for $\text{BR}(T\rightarrow St) < 0.2$. For larger values of this branching ratio, however, the combination of standard and exotic decays forbids again the existence of top partners, independently of their branching ratio into SM channels.

%

% \end{figure}
\textbf{$\mathbf{M > 1000}$ GeV:} independently of the value of $\text{BR}(T\rightarrow St)$, searches for $pp\rightarrow T\overline{T}\rightarrow t\overline{t} S S$ are no longer constraining (see fig.~\ref{fig:validation}). Likewise, standard searches loose sensitivity for $M > 1100$ GeV. For different values of $m_S$ and/or different values of the branching ratios, we refer the reader to appendix~\ref{app:software}, in which we explain how to obtain experimental bounds straightforwardly using a simple \texttt{Python} script. 

\section{Conclusions}

Heavy tops, $T$, which are predicted by different extensions of the Standard Model (SM), can decay not only into SM final states such as $Zt, Ht$ or $Wb$, but also into new particles. Among others, new scalar singlets, $S$, are good candidates. These appear, for example, in non-minimal realizations of the composite Higgs model paradigm. In light of this, we have investigated the LHC signals of top partners with sizable decay rate also into $St$. We have focused on the pair-production mode, because it depends only on the model-independent QCD interactions. We have considered ATLAS and CMS experimental analyses performed at both 8 and 13 TeV of center of mass energy. These include searches for $Ht+X$ (which are also sensitive to $St+X$ final states when $S$ decays into bottom quarks) and $Wb+X$, as well as analyses of $Zt+X$ and supersymmetry searches for pair produced stops with neutralino decays (both of which are also sensitive to $St+X$ events if $S$ is a stable particle). We have obtained the expected number of signal events passing the cuts of these analyses for arbitrary combinations of branching ratios and for different masses of $T$ and $S$. Then, we have obtained bounds resulting from a combined statistical study of all signal regions. Besides, these two procedures have been implemented in a simple code that can be found in \url{http://github.com/mikaelchala/vlqlimits}. Details about the mode of use are provided in appendix~\ref{app:software}. On the physics side, we have found that heavy tops with masses below $\sim 900$ GeV are excluded by direct searches for any value of $\text{BR}(T\rightarrow St)$, while even larger masses are forbidden for larger values of this branching ratio. All in all, we can conclude that bounds on heavy tops are not necessarily weaker if these decay also into new particles.

\section*{ Acknowledgments}%\\[-3mm]
\noindent
I am grateful to J. P. Araque and N. Castro for the useful discussions on the experimental side of this work. I would like to thank also G. Nardini and J. Santiago for useful comments on the manuscript. This work is partially supported by the Spanish MINECO under grant FPA2014-54459-P and by the Severo Ochoa Excellence Program under grant SEV-2014-0398.

\appendix

\section{Tables}\label{app:tables}
The efficiencies of the experimental analyses (see tab.~\ref{tab:effs}) for the different final states considered in this work for several values of $M$ are shown in rows $2-12$ of tab.~\ref{tab:eff100} ($m_S = 100$ GeV), tab.~\ref{tab:eff200} ($m_S = 200$ GeV), tab.~\ref{tab:eff300} ($m_S = 300$ GeV) and tab.~\ref{tab:eff400} ($m_S = 400$ GeV).
\begin{table}[t]
\begin{center}
\begin{adjustbox}{width=0.95\textwidth}
\begin{tabular}{||l|ccccccccccc||}
\hline
$M$ [GeV] & 600 & 700 & 800 & 900 & 1000 & 1100 & 1200 & 1300 & 1400 & 1500 & 1600 \\\hline
 $Zt+Zt$ & 0.004 &   0.004 &   0.006 &   0.009 &    0.010 &    0.013 &    0.015 &    0.016 &    0.017 &    0.017 &    0.017 \\
 $Zt+Ht$ & 0.002 &   0.002 &   0.003 &   0.005 &    0.008 &    0.009 &    0.012 &    0.014 &    0.015 &    0.015 &    0.015 \\
 $Zt+Wb$  &0.001 &   0.000 &   0.002 &   0.002 &    0.004 &    0.005 &    0.006 &    0.007 &    0.008 &    0.008 &    0.008 \\
 $Ht+Ht$  & 0.056 &   0.083 &   0.104 &   0.139 &    0.167 &    0.185 &    0.185 &    0.185 &    0.185 &    0.185 &    0.185 \\
 $Ht+Zt$  & 0.039 &   0.068 &   0.105 &   0.151 &    0.179 &    0.193 &    0.193 &    0.193 &    0.193 &    0.193 &    0.193 \\
 $Ht+Wb$  & 0.026 &   0.053 &   0.066 &   0.149 &    0.208 &    0.257 &    0.257 &    0.257 &    0.257 &    0.257 &    0.257 \\
 $Wb+Wb$ & 0.021 &   0.025 &   0.027 &   0.031 &    0.031 &    0.031 &    0.031 &    0.031 &    0.031 &    0.031 &    0.031 \\
 $St+Ht$ & 0.001 &   0.003 &   0.005 &   0.007 &    0.009 &    0.010 &    0.010 &    0.010 &    0.010 &    0.010 &    0.010 \\
 $St+Zt$ & 0.002 &   0.004 &   0.008 &   0.011 &    0.014 &    0.017 &    0.019 &    0.020 &    0.020 &    0.021 &    0.021 \\
 $St+Wb$ & 0.001 &   0.002 &   0.004 &   0.006 &    0.009 &    0.010 &    0.010 &    0.011 &    0.010 &    0.011 &    0.011 \\
 $St+St (1)$ & 0.009 &   0.007 &   0.007 &   0.006 &    0.006 &    0.006 &    0.007 &    0.006 &    0.006 &    0.006 &    0.006 \\
 $St+St (2)$ & 0.003 &   0.006 &   0.008 &   0.009 &    0.010 &    0.010 &    0.010 &    0.009 &    0.009 &    0.009 &    0.009 \\
 $Wb+Wb$ [pb] & 0.170 &   0.130 &   0.070 &   0.048 &    0.025 &    0.019 &    0.017 &    0.014 &    0.013 &    0.013 &    0.013 \\
 $Ht+Ht$ [pb] & 0.106 &   0.081 &   0.055 &   0.031 &    0.023 &    0.016 &    0.012 &    0.010 &    0.009 &    0.009 &    0.009 \\
\hline
\end{tabular}
\end{adjustbox}
\end{center}
\caption{\it Efficiencies of the experimental analyses (see tab.~\ref{tab:effs}) for the different final states considered in this work for $m_S = 100$ GeV and several values of $M$ (rows $2-12$). The last two rows show the upper limit on the pair production of $T$ quarks with subsequent decay into the indicated final state at 13 TeV of c.m.e. obtained from refs.~\cite{ATLAS-CONF-2016-102} and \cite{ATLAS-CONF-2016-104}, respectively. }\label{tab:eff100}
\end{table}

\begin{table}[t]
\begin{center}
\begin{adjustbox}{width=0.95\textwidth}
\begin{tabular}{||l|ccccccccccc||}
\hline
 $M$ [GeV] & 600 & 700 & 800 & 900 & 1000 & 1100 & 1200 & 1300 & 1400 & 1500 & 1600 \\\hline
 $Zt+Zt$  & 0.004 &   0.004 &   0.006 &   0.009 &    0.010 &    0.013 &    0.015 &    0.016 &    0.017 &    0.017 &    0.017 \\
$Zt+Ht$   & 0.002 &   0.002 &   0.003 &   0.005 &    0.008 &    0.009 &    0.012 &    0.014 &    0.015 &    0.015 &    0.015 \\
$Zt+Wb$   & 0.001 &   0.000 &   0.002 &   0.002 &    0.004 &    0.005 &    0.006 &    0.007 &    0.008 &    0.008 &    0.008 \\
 $Ht+Ht$  & 0.056 &   0.083 &   0.104 &   0.139 &    0.167 &    0.185 &    0.185 &    0.185 &    0.185 &    0.185 &    0.185 \\
 $Ht+Zt$  & 0.039 &   0.068 &   0.105 &   0.151 &    0.179 &    0.193 &    0.193 &    0.193 &    0.193 &    0.193 &    0.193 \\
 $Ht+Wb$  & 0.026 &   0.053 &   0.066 &   0.149 &    0.208 &    0.257 &    0.257 &    0.257 &    0.257 &    0.257 &    0.257 \\
$Wb+Wb$   & 0.021 &   0.025 &   0.027 &   0.031 &    0.031 &    0.031 &    0.031 &    0.031 &    0.031 &    0.031 &    0.031 \\
 $St+Ht$  & 0.001 &   0.003 &   0.006 &   0.008 &    0.011 &    0.012 &    0.012 &    0.014 &    0.013 &    0.014 &    0.014 \\
 $St+Zt$  & 0.002 &   0.005 &   0.009 &   0.011 &    0.017 &    0.020 &    0.022 &    0.023 &    0.023 &    0.024 &    0.024 \\
 $St+Wb$  & 0.001 &   0.002 &   0.005 &   0.007 &    0.010 &    0.010 &    0.013 &    0.013 &    0.013 &    0.012 &    0.012 \\
 $St+St(1)$  & 0.009 &   0.008 &   0.007 &   0.006 &    0.005 &    0.006 &    0.006 &    0.005 &    0.006 &    0.005 &    0.005 \\
 $St+St (2)$  & 0.004 &   0.006 &   0.009 &   0.011 &    0.012 &    0.012 &    0.012 &    0.013 &    0.013 &    0.012 &    0.012 \\
 $Wb+Wb$ [pb] & 0.170 &   0.130 &   0.070 &   0.048 &    0.025 &    0.019 &    0.017 &    0.014 &    0.013 &    0.013 &    0.013 \\
 $Ht+Ht$ [pb] & 0.106 &   0.081 &   0.055 &   0.031 &    0.023 &    0.016 &    0.012 &    0.010 &    0.009 &    0.009 &    0.009 \\
\hline
\end{tabular}
\end{adjustbox}
\end{center}
\caption{\it Same as tab.~\ref{tab:eff100} but for $m_S = 200$ GeV.}\label{tab:eff200}
\end{table}

\begin{table}[t]
\begin{center}
\begin{adjustbox}{width=0.95\textwidth}
\begin{tabular}{||l|ccccccccccc||}
\hline
$M$ [GeV] &600 & 700 & 800 & 900 & 1000 & 1100 & 1200 & 1300 & 1400 & 1500 & 1600 \\
$Zt+Zt$  &   0.004 &   0.004 &   0.006 &   0.009 &    0.010 &    0.013 &    0.015 &    0.016 &    0.017 &    0.017 &    0.017 \\
$Zt+Ht$&   0.002 &   0.002 &   0.003 &   0.005 &    0.008 &    0.009 &    0.012 &    0.014 &    0.015 &    0.015 &    0.015 \\
$Zt+Wb$&  0.001 &   0.000 &   0.002 &   0.002 &    0.004 &    0.005 &    0.006 &    0.007 &    0.008 &    0.008 &    0.008 \\
$Ht+Ht$&   0.056 &   0.083 &   0.104 &   0.139 &    0.167 &    0.185 &    0.185 &    0.185 &    0.185 &    0.185 &    0.185 \\
$Ht+Zt$&   0.039 &   0.068 &   0.105 &   0.151 &    0.179 &    0.193 &    0.193 &    0.193 &    0.193 &    0.193 &    0.193 \\
$Ht+Wb$&   0.026 &   0.053 &   0.066 &   0.149 &    0.208 &    0.257 &    0.257 &    0.257 &    0.257 &    0.257 &    0.257 \\
$Wb+Wb$&   0.021 &   0.025 &   0.027 &   0.031 &    0.031 &    0.031 &    0.031 &    0.031 &    0.031 &    0.031 &    0.031 \\
$St+Zt$&   0.002 &   0.004 &   0.007 &   0.009 &    0.015 &    0.015 &    0.015 &    0.017 &    0.016 &    0.015 &    0.015 \\
$St+Ht$&   0.003 &   0.005 &   0.010 &   0.016 &    0.019 &    0.023 &    0.024 &    0.026 &    0.025 &    0.026 &    0.026 \\
$St+Wb$&   0.001 &   0.003 &   0.006 &   0.008 &    0.011 &    0.013 &    0.014 &    0.015 &    0.015 &    0.017 &    0.017 \\
$St+St (1)$&   0.009 &   0.008 &   0.006 &   0.005 &    0.005 &    0.005 &    0.005 &    0.004 &    0.004 &    0.004 &    0.004 \\
$St+St (2)$&   0.003 &   0.007 &   0.010 &   0.012 &    0.014 &    0.015 &    0.014 &    0.015 &    0.015 &    0.014 &    0.014 \\
$Wb+Wb$ [pb]&   0.170 &   0.130 &   0.070 &   0.048 &    0.025 &    0.019 &    0.017 &    0.014 &    0.013 &    0.013 &    0.013 \\
$Ht+Ht$ [pb] &   0.106 &   0.081 &   0.055 &   0.031 &    0.023 &    0.016 &    0.012 &    0.010 &    0.009 &    0.009 &    0.009 \\
\hline
\end{tabular}
\end{adjustbox}
\end{center}
\caption{\it Same as tab.~\ref{tab:eff100} but for $m_S = 300$ GeV.}\label{tab:eff300}
\end{table}
%
%\vspace{-2.5cm}
%
\begin{table}[t]
\begin{center}
\begin{adjustbox}{width=0.95\textwidth}
\begin{tabular}{lccccccccccc}
%\hline
 & & & & & & & & & & &  \\
\end{tabular}
\end{adjustbox}
\end{center}
%\
\end{table}
\begin{table}[ht!]
\begin{center}
\begin{adjustbox}{width=0.95\textwidth}
\begin{tabular}{||l|ccccccccccc||}
\hline
$M$ [GeV] &600& 700 & 800 & 900 & 1000 & 1100 & 1200 & 1300 & 1400 & 1500 & 1600 \\
$Zt+Zt$ &   0.004 &   0.004 &   0.006 &   0.009 &    0.010 &    0.013 &    0.015 &    0.016 &    0.017 &    0.017 &    0.017 \\
$Zt+Ht$&   0.002 &   0.002 &   0.003 &   0.005 &    0.008 &    0.009 &    0.012 &    0.014 &    0.015 &    0.015 &    0.015 \\
$Zt+Wb$&  0.001 &   0.000 &   0.002 &   0.002 &    0.004 &    0.005 &    0.006 &    0.007 &    0.008 &    0.008 &    0.008 \\
$Ht+Ht$&   0.056 &   0.083 &   0.104 &   0.139 &    0.167 &    0.185 &    0.185 &    0.185 &    0.185 &    0.185 &    0.185 \\
$Ht+Zt$&   0.039 &   0.068 &   0.105 &   0.151 &    0.179 &    0.193 &    0.193 &    0.193 &    0.193 &    0.193 &    0.193 \\
$Ht+Wb$&   0.026 &   0.053 &   0.066 &   0.149 &    0.208 &    0.257 &    0.257 &    0.257 &    0.257 &    0.257 &    0.257 \\
$Wb+Wb$&   0.021 &   0.025 &   0.027 &   0.031 &    0.031 &    0.031 &    0.031 &    0.031 &    0.031 &    0.031 &    0.031 \\
$St+Zt$&   0.002 &   0.006 &   0.008 &   0.012 &    0.016 &    0.018 &    0.019 &    0.019 &    0.021 &    0.021 &    0.021 \\
$St+Ht$&   0.004 &   0.006 &   0.012 &   0.018 &    0.021 &    0.023 &    0.028 &    0.030 &    0.029 &    0.030 &    0.030 \\
$St+Wb$&   0.002 &   0.003 &   0.007 &   0.010 &    0.012 &    0.015 &    0.016 &    0.018 &    0.019 &    0.018 &    0.018 \\
$St+St (1)$&   0.008 &   0.007 &   0.006 &   0.005 &    0.005 &    0.004 &    0.004 &    0.004 &    0.004 &    0.004 &    0.004 \\
$St+St (2)$&   0.003 &   0.007 &   0.011 &   0.014 &    0.015 &    0.016 &    0.017 &    0.018 &    0.017 &    0.015 &    0.015 \\
$Ht+Ht$ [pb]&   0.170 &   0.130 &   0.070 &   0.048 &    0.025 &    0.019 &    0.017 &    0.014 &    0.013 &    0.013 &    0.013 \\
$Wb+Wb$ [pb]&   0.106 &   0.081 &   0.055 &   0.031 &    0.023 &    0.016 &    0.012 &    0.010 &    0.009 &    0.009 &    0.009 \\
\hline
\end{tabular}
\end{adjustbox}
\end{center}
\caption{\it Same as tab.~\ref{tab:eff100} but for $m_S = 400$ GeV.}\label{tab:eff400}
\end{table}
 
\section{Software}\label{app:software}
 A small \texttt{Python} script can be found in \url{http://github.com/mikaelchala/vlqlimits}. The usage is extremely simple. First, one needs to edit the file \texttt{input.dat} (the name can be of course changed). It reads:
 \footnotesize{
 \begin{verbatim}
% THIS IS A COMMENT
% FORMAT: M, MS, BR(T->WB), BR(T->HT)+BR(T->ST, S->BB), BR(T->ZT), BR(T->ST, S -> INV)
  735.    210.    0.5     0.1     0.1     0.3
 \end{verbatim}}
 \normalsize
The non-commented line contains an example point. More points, corresponding to other possible quarks present in the spectrum, can be added (interference effects are disregarded). We recall that the branching ratios do not have to add to 1. This would reflect the existence of further decays for which there is no experimental information (for example, decays into very soft particles, which are hard to detect). Given this, one can compute whether the set of heavy tops indicated in the file is excluded at the 95 \% C.L. or not. This can be done in two ways:
 \begin{itemize}
  \item \texttt{./code\_indp.py input.dat}: returns \texttt{1} if the number of predicted signal events in at least one signal region is above that quoted in tab.~\ref{tab:effs}, or if the cross sections in $Wb+Wb$ or $Ht+Ht$ exceed the ones reported by the corresponding analyses at 13 TeV of c.m.e.~\cite{ATLAS-CONF-2016-102,ATLAS-CONF-2016-104}. It returns \texttt{0} otherwise.
  \item \texttt{./code\_comb.py input.dat}: returns \texttt{1} if the CL$_s$ computed considering all signal regions in tab.~\ref{tab:effs} is below $0.05$. Or if the cross sections in $Wb+Wb$ or $Ht+Ht$ exceed the ones reported by the corresponding analyses at 13 TeV of c.m.e.~\cite{ATLAS-CONF-2016-102,ATLAS-CONF-2016-104}. It returns \texttt{0} otherwise. This code can be only run if \texttt{PyROOT} is properly installed.
 \end{itemize}
 In both cases, the \texttt{NumPy} and the \texttt{SciPy} libraries are mandatory. The cross sections and efficiencies for masses not considered in the Monte Carlo simulations are linearly interpolated from those included in the hidden file \texttt{.tables}. More sophisticated tools, that include also heavy fermions with different quantum numbers, are also publicly available (see for example \texttt{XQCAT}~\cite{Barducci:2014ila, Barducci:2014gna}). However, exotic decays and signals at 13 TeV are generally not considered. In this respect, this code complements previous works.
\noindent

\clearpage
\bibliographystyle{JHEP}
\bibliography{notes}{}

\providecommand{\href}[2]{#2}\begingroup\raggedright\begin{thebibliography}{10}

\bibitem{AguilarSaavedra:2005pv}
J.~A. Aguilar-Saavedra, \emph{{Pair production of heavy Q = 2/3 singlets at
  LHC}}, \href{http://dx.doi.org/10.1016/j.physletb.2005.08.062,
  10.1016/j.physletb.2005.12.013}{\emph{Phys. Lett.} {\bf B625} (2005)
  234--244}, [\href{https://arxiv.org/abs/hep-ph/0506187}{{\tt
  hep-ph/0506187}}].

\bibitem{AguilarSaavedra:2009es}
J.~A. Aguilar-Saavedra, \emph{{Identifying top partners at LHC}},
  \href{http://dx.doi.org/10.1088/1126-6708/2009/11/030}{\emph{JHEP} {\bf 11}
  (2009) 030}, [\href{https://arxiv.org/abs/0907.3155}{{\tt 0907.3155}}].

\bibitem{Aguilar-Saavedra:2013qpa}
J.~A. Aguilar-Saavedra, R.~Benbrik, S.~Heinemeyer and M.~Pérez-Victoria,
  \emph{{Handbook of vectorlike quarks: Mixing and single production}},
  \href{http://dx.doi.org/10.1103/PhysRevD.88.094010}{\emph{Phys. Rev.} {\bf
  D88} (2013) 094010}, [\href{https://arxiv.org/abs/1306.0572}{{\tt
  1306.0572}}].

\bibitem{Kaplan:1983fs}
D.~B. Kaplan and H.~Georgi, \emph{{SU(2) x U(1) Breaking by Vacuum
  Misalignment}},
  \href{http://dx.doi.org/10.1016/0370-2693(84)91177-8}{\emph{Phys. Lett.} {\bf
  B136} (1984) 183--186}.

\bibitem{Kaplan:1983sm}
D.~B. Kaplan, H.~Georgi and S.~Dimopoulos, \emph{{Composite Higgs Scalars}},
  \href{http://dx.doi.org/10.1016/0370-2693(84)91178-X}{\emph{Phys. Lett.} {\bf
  B136} (1984) 187--190}.

\bibitem{Dimopoulos:1981xc}
S.~Dimopoulos and J.~Preskill, \emph{{Massless Composites With Massive
  Constituents}},
  \href{http://dx.doi.org/10.1016/0550-3213(82)90345-5}{\emph{Nucl. Phys.} {\bf
  B199} (1982) 206--222}.

\bibitem{Kaplan:1991dc}
D.~B. Kaplan, \emph{{Flavor at SSC energies: A New mechanism for dynamically
  generated fermion masses}},
  \href{http://dx.doi.org/10.1016/S0550-3213(05)80021-5}{\emph{Nucl. Phys.}
  {\bf B365} (1991) 259--278}.

\bibitem{Agashe:2004rs}
K.~Agashe, R.~Contino and A.~Pomarol, \emph{{The Minimal composite Higgs
  model}}, \href{http://dx.doi.org/10.1016/j.nuclphysb.2005.04.035}{\emph{Nucl.
  Phys.} {\bf B719} (2005) 165--187},
  [\href{https://arxiv.org/abs/hep-ph/0412089}{{\tt hep-ph/0412089}}].

\bibitem{Gripaios:2009pe}
B.~Gripaios, A.~Pomarol, F.~Riva and J.~Serra, \emph{{Beyond the Minimal
  Composite Higgs Model}},
  \href{http://dx.doi.org/10.1088/1126-6708/2009/04/070}{\emph{JHEP} {\bf 04}
  (2009) 070}, [\href{https://arxiv.org/abs/0902.1483}{{\tt 0902.1483}}].

\bibitem{Chala:2016ykx}
M.~Chala, G.~Nardini and I.~Sobolev, \emph{{Unified explanation for dark matter
  and electroweak baryogenesis with direct detection and gravitational wave
  signatures}}, \href{http://dx.doi.org/10.1103/PhysRevD.94.055006}{\emph{Phys.
  Rev.} {\bf D94} (2016) 055006}, [\href{https://arxiv.org/abs/1605.08663}{{\tt
  1605.08663}}].

\bibitem{Chala:2012af}
M.~Chala, \emph{{$h \rightarrow \gamma\gamma$ excess and Dark Matter from
  Composite Higgs Models}},
  \href{http://dx.doi.org/10.1007/JHEP01(2013)122}{\emph{JHEP} {\bf 01} (2013)
  122}, [\href{https://arxiv.org/abs/1210.6208}{{\tt 1210.6208}}].

\bibitem{Sanz:2015sua}
V.~Sanz and J.~Setford, \emph{{Composite Higgses with seesaw EWSB}},
  \href{http://dx.doi.org/10.1007/JHEP12(2015)154}{\emph{JHEP} {\bf 12} (2015)
  154}, [\href{https://arxiv.org/abs/1508.06133}{{\tt 1508.06133}}].

\bibitem{Vecchi:2013bja}
L.~Vecchi, \emph{{The Natural Composite Higgs}},
  \href{https://arxiv.org/abs/1304.4579}{{\tt 1304.4579}}.

\bibitem{Frigerio:2012uc}
M.~Frigerio, A.~Pomarol, F.~Riva and A.~Urbano, \emph{{Composite Scalar Dark
  Matter}}, \href{http://dx.doi.org/10.1007/JHEP07(2012)015}{\emph{JHEP} {\bf
  07} (2012) 015}, [\href{https://arxiv.org/abs/1204.2808}{{\tt 1204.2808}}].

\bibitem{Mrazek:2011iu}
J.~Mrazek, A.~Pomarol, R.~Rattazzi, M.~Redi, J.~Serra and A.~Wulzer, \emph{{The
  Other Natural Two Higgs Doublet Model}},
  \href{http://dx.doi.org/10.1016/j.nuclphysb.2011.07.008}{\emph{Nucl. Phys.}
  {\bf B853} (2011) 1--48}, [\href{https://arxiv.org/abs/1105.5403}{{\tt
  1105.5403}}].

\bibitem{Fonseca:2015gva}
N.~Fonseca, R.~Zukanovich~Funchal, A.~Lessa and L.~Lopez-Honorez, \emph{{Dark
  Matter Constraints on Composite Higgs Models}},
  \href{http://dx.doi.org/10.1007/JHEP06(2015)154}{\emph{JHEP} {\bf 06} (2015)
  154}, [\href{https://arxiv.org/abs/1501.05957}{{\tt 1501.05957}}].

\bibitem{Ma:2017vzm}
Y.~Wu, B.~Zhang, T.~Ma and G.~Cacciapaglia, \emph{{Composite Dark Matter and
  Higgs}},  \href{https://arxiv.org/abs/1703.06903}{{\tt 1703.06903}}.

\bibitem{Ballesteros:2017xeg}
G.~Ballesteros, A.~Carmona and M.~Chala, \emph{{Exceptional Composite Dark
  Matter}},  \href{https://arxiv.org/abs/1704.07388}{{\tt 1704.07388}}.

\bibitem{Espinosa:2011eu}
J.~R. Espinosa, B.~Gripaios, T.~Konstandin and F.~Riva, \emph{{Electroweak
  Baryogenesis in Non-minimal Composite Higgs Models}},
  \href{http://dx.doi.org/10.1088/1475-7516/2012/01/012}{\emph{JCAP} {\bf 1201}
  (2012) 012}, [\href{https://arxiv.org/abs/1110.2876}{{\tt 1110.2876}}].

\bibitem{Caracciolo:2012je}
F.~Caracciolo, A.~Parolini and M.~Serone, \emph{{UV Completions of Composite
  Higgs Models with Partial Compositeness}},
  \href{http://dx.doi.org/10.1007/JHEP02(2013)066}{\emph{JHEP} {\bf 02} (2013)
  066}, [\href{https://arxiv.org/abs/1211.7290}{{\tt 1211.7290}}].

\bibitem{Barnard:2013zea}
J.~Barnard, T.~Gherghetta and T.~S. Ray, \emph{{UV descriptions of composite
  Higgs models without elementary scalars}},
  \href{http://dx.doi.org/10.1007/JHEP02(2014)002}{\emph{JHEP} {\bf 02} (2014)
  002}, [\href{https://arxiv.org/abs/1311.6562}{{\tt 1311.6562}}].

\bibitem{Ferretti:2014qta}
G.~Ferretti, \emph{{UV Completions of Partial Compositeness: The Case for a
  SU(4) Gauge Group}},
  \href{http://dx.doi.org/10.1007/JHEP06(2014)142}{\emph{JHEP} {\bf 06} (2014)
  142}, [\href{https://arxiv.org/abs/1404.7137}{{\tt 1404.7137}}].

\bibitem{Vecchi:2015fma}
L.~Vecchi, \emph{{A dangerous irrelevant UV-completion of the composite
  Higgs}}, \href{http://dx.doi.org/10.1007/JHEP02(2017)094}{\emph{JHEP} {\bf
  02} (2017) 094}, [\href{https://arxiv.org/abs/1506.00623}{{\tt 1506.00623}}].

\bibitem{Ma:2015gra}
T.~Ma and G.~Cacciapaglia, \emph{{Fundamental Composite 2HDM: SU(N) with 4
  flavours}}, \href{http://dx.doi.org/10.1007/JHEP03(2016)211}{\emph{JHEP} {\bf
  03} (2016) 211}, [\href{https://arxiv.org/abs/1508.07014}{{\tt 1508.07014}}].

\bibitem{Ferretti:2016upr}
G.~Ferretti, \emph{{Gauge theories of Partial Compositeness: Scenarios for
  Run-II of the LHC}},
  \href{http://dx.doi.org/10.1007/JHEP06(2016)107}{\emph{JHEP} {\bf 06} (2016)
  107}, [\href{https://arxiv.org/abs/1604.06467}{{\tt 1604.06467}}].

\bibitem{Serra:2015xfa}
J.~Serra, \emph{{Beyond the Minimal Top Partner Decay}},
  \href{http://dx.doi.org/10.1007/JHEP09(2015)176}{\emph{JHEP} {\bf 09} (2015)
  176}, [\href{https://arxiv.org/abs/1506.05110}{{\tt 1506.05110}}].

\bibitem{Anandakrishnan:2015yfa}
A.~Anandakrishnan, J.~H. Collins, M.~Farina, E.~Kuflik and M.~Perelstein,
  \emph{{Odd Top Partners at the LHC}},
  \href{http://dx.doi.org/10.1103/PhysRevD.93.075009}{\emph{Phys. Rev.} {\bf
  D93} (2016) 075009}, [\href{https://arxiv.org/abs/1506.05130}{{\tt
  1506.05130}}].

\bibitem{Cacciapaglia:2015eqa}
G.~Cacciapaglia, H.~Cai, A.~Deandrea, T.~Flacke, S.~J. Lee and A.~Parolini,
  \emph{{Composite scalars at the LHC: the Higgs, the Sextet and the Octet}},
  \href{http://dx.doi.org/10.1007/JHEP11(2015)201}{\emph{JHEP} {\bf 11} (2015)
  201}, [\href{https://arxiv.org/abs/1507.02283}{{\tt 1507.02283}}].

\bibitem{Fan:2015sza}
J.~Fan, S.~M. Koushiappas and G.~Landsberg, \emph{{Pseudoscalar Portal Dark
  Matter and New Signatures of Vector-like Fermions}},
  \href{http://dx.doi.org/10.1007/JHEP01(2016)111}{\emph{JHEP} {\bf 01} (2016)
  111}, [\href{https://arxiv.org/abs/1507.06993}{{\tt 1507.06993}}].

\bibitem{Banerjee:2016wls}
S.~Banerjee, D.~Barducci, G.~Bélanger and C.~Delaunay, \emph{{Implications of
  a High-Mass Diphoton Resonance for Heavy Quark Searches}},
  \href{http://dx.doi.org/10.1007/JHEP11(2016)154}{\emph{JHEP} {\bf 11} (2016)
  154}, [\href{https://arxiv.org/abs/1606.09013}{{\tt 1606.09013}}].

\bibitem{Niehoff:2016zso}
C.~Niehoff, P.~Stangl and D.~M. Straub, \emph{{Electroweak symmetry breaking
  and collider signatures in the next-to-minimal composite Higgs model}},
  \href{https://arxiv.org/abs/1611.09356}{{\tt 1611.09356}}.

\bibitem{Barducci:2014ila}
D.~Barducci, A.~Belyaev, M.~Buchkremer, G.~Cacciapaglia, A.~Deandrea,
  S.~De~Curtis et~al., \emph{{Framework for Model Independent Analyses of
  Multiple Extra Quark Scenarios}},
  \href{http://dx.doi.org/10.1007/JHEP12(2014)080}{\emph{JHEP} {\bf 12} (2014)
  080}, [\href{https://arxiv.org/abs/1405.0737}{{\tt 1405.0737}}].

\bibitem{Aad:2015kqa}
{\scshape ATLAS} collaboration, G.~Aad et~al., \emph{{Search for production of
  vector-like quark pairs and of four top quarks in the lepton-plus-jets final
  state in $pp$ collisions at $\sqrt{s}=8$ TeV with the ATLAS detector}},
  \href{http://dx.doi.org/10.1007/JHEP08(2015)105}{\emph{JHEP} {\bf 08} (2015)
  105}, [\href{https://arxiv.org/abs/1505.04306}{{\tt 1505.04306}}].

\bibitem{ATLAS-CONF-2016-102}
{\scshape ATLAS Collaboration} collaboration, \emph{{Search for pair production
  of heavy vector-like quarks decaying to high-$p_T$ $W$ bosons and b quarks in
  the lepton-plus-jets final state in pp collisions at $\sqrt{s}$=13 TeV with
  the ATLAS detector}},  Tech. Rep. ATLAS-CONF-2016-102, CERN, Geneva, Sep,
  2016.

\bibitem{ATLAS-CONF-2016-104}
{\scshape ATLAS Collaboration} collaboration, \emph{{Search for new phenomena
  in $t\bar{t}$ final states with additional heavy-flavour jets in $pp$
  collisions at $\sqrt{s}=13$ TeV with the ATLAS detector}},  Tech. Rep.
  ATLAS-CONF-2016-104, CERN, Geneva, Sep, 2016.

\bibitem{ATLAS-CONF-2017-015}
{\scshape ATLAS Collaboration} collaboration, \emph{{Search for pair production
  of vector-like top quarks in events with one lepton and an invisibly decaying
  Z boson in $\sqrt{s} = 13$ TeV pp collisions at the ATLAS detector}},  Tech.
  Rep. ATLAS-CONF-2017-015, CERN, Geneva, Mar, 2017.

\bibitem{CMS:2016hxa}
{\scshape CMS} collaboration, C.~Collaboration, \emph{{Search for direct top
  squark pair production in the fully hadronic final state in proton-proton
  collisions at sqrt(s) = 13 TeV corresponding to an integrated luminosity of
  12.9/fb}}, .

\bibitem{Contino:2006qr}
R.~Contino, L.~Da~Rold and A.~Pomarol, \emph{{Light custodians in natural
  composite Higgs models}},
  \href{http://dx.doi.org/10.1103/PhysRevD.75.055014}{\emph{Phys. Rev.} {\bf
  D75} (2007) 055014}, [\href{https://arxiv.org/abs/hep-ph/0612048}{{\tt
  hep-ph/0612048}}].

\bibitem{Matsedonskyi:2012ym}
O.~Matsedonskyi, G.~Panico and A.~Wulzer, \emph{{Light Top Partners for a Light
  Composite Higgs}},
  \href{http://dx.doi.org/10.1007/JHEP01(2013)164}{\emph{JHEP} {\bf 01} (2013)
  164}, [\href{https://arxiv.org/abs/1204.6333}{{\tt 1204.6333}}].

\bibitem{Redi:2012ha}
M.~Redi and A.~Tesi, \emph{Implications of a light {H}iggs in composite
  models}, \href{http://dx.doi.org/10.1007/JHEP10(2012)166}{\emph{JHEP} {\bf
  1210} (2012) 166}, [\href{https://arxiv.org/abs/1205.0232}{{\tt 1205.0232}}].

\bibitem{Marzocca:2012zn}
D.~Marzocca, M.~Serone and J.~Shu, \emph{{General Composite Higgs Models}},
  \href{http://dx.doi.org/10.1007/JHEP08(2012)013}{\emph{JHEP} {\bf 08} (2012)
  013}, [\href{https://arxiv.org/abs/1205.0770}{{\tt 1205.0770}}].

\bibitem{Pomarol:2012qf}
A.~Pomarol and F.~Riva, \emph{The composite {H}iggs and light resonance
  connection}, \href{http://dx.doi.org/10.1007/JHEP08(2012)135}{\emph{JHEP}
  {\bf 1208} (2012) 135}, [\href{https://arxiv.org/abs/1205.6434}{{\tt
  1205.6434}}].

\bibitem{Panico:2012uw}
G.~Panico, M.~Redi, A.~Tesi and A.~Wulzer, \emph{{On the Tuning and the Mass of
  the Composite Higgs}},
  \href{http://dx.doi.org/10.1007/JHEP03(2013)051}{\emph{JHEP} {\bf 03} (2013)
  051}, [\href{https://arxiv.org/abs/1210.7114}{{\tt 1210.7114}}].

\bibitem{Chala:2017sjk}
M.~Chala, G.~Durieux, C.~Grojean, L.~de~Lima and O.~Matsedonskyi,
  \emph{{Minimally extended SILH}},
  \href{https://arxiv.org/abs/1703.10624}{{\tt 1703.10624}}.

\bibitem{Khachatryan:2016vau}
{\scshape ATLAS, CMS} collaboration, G.~Aad et~al., \emph{{Measurements of the
  Higgs boson production and decay rates and constraints on its couplings from
  a combined ATLAS and CMS analysis of the LHC pp collision data at $
  \sqrt{s}=7 $ and 8 TeV}},
  \href{http://dx.doi.org/10.1007/JHEP08(2016)045}{\emph{JHEP} {\bf 08} (2016)
  045}, [\href{https://arxiv.org/abs/1606.02266}{{\tt 1606.02266}}].

\bibitem{Ghosh:2015wiz}
D.~Ghosh, M.~Salvarezza and F.~Senia, \emph{{Extending the Analysis of
  Electroweak Precision Constraints in Composite Higgs Models}},
  \href{http://dx.doi.org/10.1016/j.nuclphysb.2016.11.013}{\emph{Nucl. Phys.}
  {\bf B914} (2017) 346--387}, [\href{https://arxiv.org/abs/1511.08235}{{\tt
  1511.08235}}].

\bibitem{Chala:2014mma}
M.~Chala, J.~Juknevich, G.~Perez and J.~Santiago, \emph{{The Elusive Gluon}},
  \href{http://dx.doi.org/10.1007/JHEP01(2015)092}{\emph{JHEP} {\bf 01} (2015)
  092}, [\href{https://arxiv.org/abs/1411.1771}{{\tt 1411.1771}}].

\bibitem{Araque:2015cna}
J.~P. Araque, N.~F. Castro and J.~Santiago, \emph{{Interpretation of
  Vector-like Quark Searches: Heavy Gluons in Composite Higgs Models}},
  \href{http://dx.doi.org/10.1007/JHEP11(2015)120}{\emph{JHEP} {\bf 11} (2015)
  120}, [\href{https://arxiv.org/abs/1507.05628}{{\tt 1507.05628}}].

\bibitem{Azatov:2015xqa}
A.~Azatov, D.~Chowdhury, D.~Ghosh and T.~S. Ray, \emph{{Same sign di-lepton
  candles of the composite gluons}},
  \href{http://dx.doi.org/10.1007/JHEP08(2015)140}{\emph{JHEP} {\bf 08} (2015)
  140}, [\href{https://arxiv.org/abs/1505.01506}{{\tt 1505.01506}}].

\bibitem{Matsedonskyi:2014mna}
O.~Matsedonskyi, G.~Panico and A.~Wulzer, \emph{{On the Interpretation of Top
  Partners Searches}},
  \href{http://dx.doi.org/10.1007/JHEP12(2014)097}{\emph{JHEP} {\bf 12} (2014)
  097}, [\href{https://arxiv.org/abs/1409.0100}{{\tt 1409.0100}}].

\bibitem{Matsedonskyi:2015dns}
O.~Matsedonskyi, G.~Panico and A.~Wulzer, \emph{{Top Partners Searches and
  Composite Higgs Models}},
  \href{http://dx.doi.org/10.1007/JHEP04(2016)003}{\emph{JHEP} {\bf 04} (2016)
  003}, [\href{https://arxiv.org/abs/1512.04356}{{\tt 1512.04356}}].

\bibitem{Araque:2016jrb}
{\scshape ATLAS} collaboration, J.~P. Araque, \emph{{Overview of the
  vector-like quark searches with the LHC data collected by the ATLAS
  detector}},  in \emph{{9th International Workshop on Top Quark Physics (TOP
  2016) Olomouc, Czech Republic, September 19-23, 2016}}, 2016.
\newblock \href{https://arxiv.org/abs/1611.09056}{{\tt 1611.09056}}.

\bibitem{Read:2002hq}
A.~L. Read, \emph{{Presentation of search results: The CL(s) technique}},
  \href{http://dx.doi.org/10.1088/0954-3899/28/10/313}{\emph{J. Phys.} {\bf
  G28} (2002) 2693--2704}.

\bibitem{Brun:1997pa}
R.~Brun and F.~Rademakers, \emph{{ROOT: An object oriented data analysis
  framework}},
  \href{http://dx.doi.org/10.1016/S0168-9002(97)00048-X}{\emph{Nucl. Instrum.
  Meth.} {\bf A389} (1997) 81--86}.

\bibitem{Alwall:2014hca}
J.~Alwall, R.~Frederix, S.~Frixione, V.~Hirschi, F.~Maltoni, O.~Mattelaer
  et~al., \emph{{The automated computation of tree-level and next-to-leading
  order differential cross sections, and their matching to parton shower
  simulations}}, \href{http://dx.doi.org/10.1007/JHEP07(2014)079}{\emph{JHEP}
  {\bf 07} (2014) 079}, [\href{https://arxiv.org/abs/1405.0301}{{\tt
  1405.0301}}].

\bibitem{Sjostrand:2006za}
T.~Sjostrand, S.~Mrenna and P.~Z. Skands, \emph{{PYTHIA 6.4 Physics and
  Manual}}, \href{http://dx.doi.org/10.1088/1126-6708/2006/05/026}{\emph{JHEP}
  {\bf 05} (2006) 026}, [\href{https://arxiv.org/abs/hep-ph/0603175}{{\tt
  hep-ph/0603175}}].

\bibitem{Conte:2012fm}
E.~Conte, B.~Fuks and G.~Serret, \emph{{MadAnalysis 5, A User-Friendly
  Framework for Collider Phenomenology}},
  \href{http://dx.doi.org/10.1016/j.cpc.2012.09.009}{\emph{Comput. Phys.
  Commun.} {\bf 184} (2013) 222--256},
  [\href{https://arxiv.org/abs/1206.1599}{{\tt 1206.1599}}].

\bibitem{Conte:2014zja}
E.~Conte, B.~Dumont, B.~Fuks and C.~Wymant, \emph{{Designing and recasting LHC
  analyses with MadAnalysis 5}},
  \href{http://dx.doi.org/10.1140/epjc/s10052-014-3103-0}{\emph{Eur. Phys. J.}
  {\bf C74} (2014) 3103}, [\href{https://arxiv.org/abs/1405.3982}{{\tt
  1405.3982}}].

\bibitem{Cacciari:2011ma}
M.~Cacciari, G.~P. Salam and G.~Soyez, \emph{{FastJet User Manual}},
  \href{http://dx.doi.org/10.1140/epjc/s10052-012-1896-2}{\emph{Eur. Phys. J.}
  {\bf C72} (2012) 1896}, [\href{https://arxiv.org/abs/1111.6097}{{\tt
  1111.6097}}].

\bibitem{ATLAS:2016jaa}
{\scshape ATLAS} collaboration, T.~A. collaboration, \emph{{Search for the
  Supersymmetric Partner of the Top Quark in the Jets+Emiss Final State at
  sqrt(s) = 13 TeV}}, .

\bibitem{Cacciari:2008gp}
M.~Cacciari, G.~P. Salam and G.~Soyez, \emph{{The Anti-k(t) jet clustering
  algorithm}},
  \href{http://dx.doi.org/10.1088/1126-6708/2008/04/063}{\emph{JHEP} {\bf 04}
  (2008) 063}, [\href{https://arxiv.org/abs/0802.1189}{{\tt 0802.1189}}].

\bibitem{Dokshitzer:1997in}
Y.~L. Dokshitzer, G.~D. Leder, S.~Moretti and B.~R. Webber, \emph{{Better jet
  clustering algorithms}},
  \href{http://dx.doi.org/10.1088/1126-6708/1997/08/001}{\emph{JHEP} {\bf 08}
  (1997) 001}, [\href{https://arxiv.org/abs/hep-ph/9707323}{{\tt
  hep-ph/9707323}}].

\bibitem{Barducci:2014gna}
D.~Barducci, A.~Belyaev, M.~Buchkremer, J.~Marrouche, S.~Moretti and
  L.~Panizzi, \emph{{XQCAT: eXtra Quark Combined Analysis Tool}},
  \href{http://dx.doi.org/10.1016/j.cpc.2015.08.016}{\emph{Comput. Phys.
  Commun.} {\bf 197} (2015) 263--275},
  [\href{https://arxiv.org/abs/1409.3116}{{\tt 1409.3116}}].

\end{thebibliography}\endgroup

\end{document}